\documentclass[format=acmsmall, review=false, screen=true]{acmart}

\usepackage{booktabs} % For formal tables
\usepackage{makecell}
\usepackage{lipsum}
\usepackage[utf8]{inputenc}
\usepackage{array}
\usepackage{wrapfig}
\usepackage{multirow}
\usepackage{tabularx}
\usepackage{pifont}
\usepackage{lineno}
\usepackage{balance}
\usepackage{xcolor,pifont}
\usepackage{caption}
\usepackage{soul}

\usepackage[linesnumbered, ruled,vlined]{algorithm2e}
\RequirePackage{pdflscape}
\usepackage{flushend}
\usepackage{multicol}
\usepackage{placeins}
\usepackage{booktabs}
\usepackage{indentfirst}
\usepackage[figuresright]{rotating}
\usepackage{fancybox}
\usepackage{fontawesome}
\usepackage{threeparttable}
\usepackage{mathtools}
\usepackage{MnSymbol,wasysym}
\usepackage{rotating}
\usepackage{adjustbox}
\usepackage{graphicx}
\usepackage{tcolorbox}
\newtcolorbox{mybox}[2][]
{colback = white, colframe = black, fonttitle = \bfseries,
    colbacktitle = gray, enhanced,
    attach boxed title to top left={yshift=-3mm, xshift=3mm},
    title=#2, #1}

\usepackage{framed}
\definecolor{Gray}{gray}{0.9}
\definecolor{shadecolor}{gray}{0.95}
\usepackage{tikz}
\usetikzlibrary{trees,positioning,shapes,shadows,arrows}
\tikzset{
  basic/.style  = {draw, text width=2cm, drop shadow, font=\sffamily, rectangle},
  root/.style   = {basic, rounded corners=2pt, thin, align=center, fill=white},
  level-2/.style = {basic, rounded corners=6pt, thin,align=center, fill=white, text width=3cm},
  level-3/.style = {basic, thin, align=center, fill=white, text width=1.8cm}
}
\newcommand{\todo}[1]{}
\renewcommand{\todo}[1]{{\color{red} TODO: {#1}}}
\newcommand{\sdgc}{self-declared AI-generated code}

\newcommand{\aigc}{AI-generated code}

\newcommand{\snl}{snippet-level}

\newcommand{\fl}{file-level}

\newcommand{\sdc}{self-declaration comments}

\begin{document}
\title{On Developers' Self-Declaration of AI-Generated Code: An Analysis of Practices}
%\title{How and Why Developers Self-Declare their AI-Generated Code? An Analysis of Practices}

\author{Syed Mohammad Kashif}
\affiliation{%
  \institution{School of Computer Science, Wuhan University}
  % \streetaddress{P.O. Box 1212}
  \city{Wuhan}
  \country{China}
}
\email{smkashif145@gmail.com}

\author{Peng Liang}
%\authornote{Corresponding author}
\affiliation{%
  \institution{School of Computer Science, Wuhan University}
  % \streetaddress{P.O. Box 1212}
  \city{Wuhan}
  \country{China}
  }
\email{liangp@whu.edu.cn}

\author{Amjed Tahir}
\affiliation{%
  \institution{School of Mathematical and Computational Sciences, Massey University}
  % \streetaddress{P.O. Box 1212}
  \city{Palmerston North}
  \country{New Zealand}
  }
\email{a.tahir@massey.ac.nz}

%%
%% These commands are for a JOURNAL article.
% \acmJournal{TOSEM}
% \acmVolume{0}
% \acmNumber{0}
% \acmArticle{0}
% \acmMonth{0}

\setcopyright{acmlicensed}
\acmJournal{TOSEM}
\acmYear{2025} \acmVolume{1} \acmNumber{1} \acmArticle{1} \acmMonth{1}\acmDOI{10.1145/3771937}

\renewcommand{\shortauthors}{Kashif et al.}

\begin{abstract}
AI code generation tools have gained significant popularity among developers, who use them to assist in software development due to their capability to generate code. Existing studies mainly explored the quality, e.g., correctness and security, of \aigc{}, while in real-world software development, the prerequisite is to distinguish \aigc{} from human-written code, which emphasizes the need to explicitly declare \aigc{} by developers. To this end, this study intends to understand the ways developers use to self-declare \aigc{} and explore the reasons why developers choose to self-declare or not. We conducted a mixed-methods study consisting of two phases. In the first phase, we mined GitHub repositories and collected 613 instances of \aigc{} snippets. In the second phase, we conducted a follow-up practitioners’ survey, which received 111 valid responses. Our research revealed the practices followed by developers to self-declare \aigc{}. Most practitioners (76.6\%) \textit{always} or \textit{sometimes} self-declare \aigc{}. In contrast, other practitioners (23.4\%) noted that they \textit{never} self-declare \aigc{}. The reasons for self-declaring \aigc{} include the need to track and monitor the code for future review and debugging, and ethical considerations. The reasons for not self-declaring \aigc{} include extensive modifications to \aigc{} and the developers' perception that self-declaration is an unnecessary activity. We finally provided guidelines for practitioners to self-declare \aigc{}, addressing ethical and code quality concerns. 
\end{abstract}






\maketitle

\begin{sloppypar}

\section{Introduction} \label{introduction}
Artificial Intelligence (AI) techniques, especially those employing Machine Learning (ML) models and Large Language Models (LLMs), are revolutionizing software engineering practices. There has been increased adoption of these models to develop more accurate, faster, and scalable tools that assist software engineers in their development tasks~\cite{yellin2023premature, hou2024llm4se} (e.g., AI-assisted coding~\cite{UsabilityOfAIProgramming}). %Those models have impac ted the whole software development life cycle, from requirements elicitation and design to deployment and maintenance \cite{hou2024large,khomh2023intelligent}. 
One area that has been largely transformed with the advancement in LLMs is automatic code generation \cite{chen2021evaluating, hou2024llm4se}. Developers increasingly rely on AI code generation tools in their written code \cite{AItoolsSurvey2024,AIRewriteCoding2023}. The most recent Stack Overflow Developer Survey (2024) shows that there has been a significant adoption of code generation tools in practice \cite{SO-survey2024, SO-survey2025}. This is partially due to the widely available coding-specific tools (e.g., GitHub Copilot \cite{Copilot} and CodeWhisperer \cite{CodeWhisperer}) and general-purpose LLMs (e.g., ChatGPT \cite{ChatGPT} and Gemini \cite{Gemini}) that can generate a large amount of code with little human intervention, faster with high efficiency \cite{dakhel2023github,tian2023chatgpt}. %These tools can generate human-like code fast and with high efficiency \cite{dakhel2023github,tian2023chatgpt}. 
%AI code generation tools have also been shown to have the potential to free practitioners from coding tasks~\cite{NewProgrammingPractice}.
% Existing studies explored the usability of AI code generation tools in software development. For instance, a study specifically on Copilot indicates that developers use AI tools in development because it helps suggest a good starting point to complete their tasks instead of spending time online searching for solutions~\cite{vaithilingam2022expectation}. It is also evident from the literature that developers are using these AI tools mainly to recall syntax, optimize key-stroke, and complete development tasks fast to save time~\cite{UsabilityOfAIProgramming}.

Despite their usefulness and high potential \cite{RocksCodingNotDevelopment,UsabilityOfAIProgramming}, previous studies have shown that there are many quality \cite{yeticstiren2023evaluating,liu2024no}, correctness \cite{tambon2025bugs,vatai2024tadashi,liu2024your}, security \cite{pearce2022asleep,majdinasab2024assessing,fu2023security} and ethical ~\cite{stalnaker2024developer} challenges associated with \aigc{} (i.e., code that is primarily generated with AI coding assistance tools). These challenges stress the need to distinguish \aigc{} from other human-written code \cite{xu2024distinguishing, idialu2024whodunit} for code review and quality assurance purposes \cite{xu2024distinguishing}. To this end, we seek to study whether developers self-declare \aigc{}, explore the characteristics of self-declaration comments, and explore how and why developers self-declare (or not) \aigc{}. We derive our definition of \textbf{\sdgc{}} from the well-established definition of \textit{self-admitted technical debt} \cite{potdar2014exploratory}: \textit{a \sdgc{}} is code that has been explicitly acknowledged in the code comments by the developer as automatically generated by AI tools.

\begin{figure}[h]
    \centering
    \includegraphics[width=0.8\linewidth]{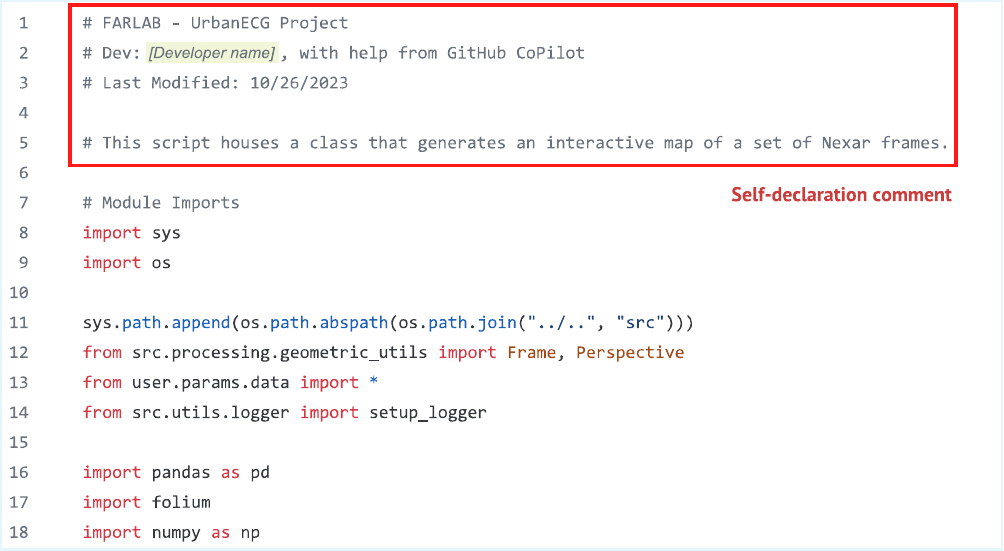}
    \caption{\textcolor{black}{An example of \sdc{} of \aigc{} (\href{\detokenize{https://github.com/pendulating/urban-fingerprinting/blob/35e874a37131bdc7830341811f22339c8216052e/src/visualization/inspection_map.py\#L2}}{urban-fingerprinting})}}
    \label{fig:comments}
    \vspace{-1em}
\end{figure}

\textbf{A motivating example:} We observed in many GitHub repositories that developers often include code comments indicating that certain lines of code or specific files were auto-generated using AI code generation tools. The study of Fu \textit{et al.} collected a dataset of such code comments to investigate security weaknesses in \aigc{}~\cite{fu2023security}. Other studies have also used a similar approach of collecting \aigc{} from GitHub repositories~\cite{yu2024large, tambon2025bugs}. The authors used code comments from the GitHub repositories to identify \aigc{}. We refer to such comments as self-declaration comments. A sample self-declaration comment from an \aigc{} snippet is illustrated in Figure \ref{fig:comments}. In this example, a developer acknowledges that the code file was written with the help of GitHub Copilot, an AI code generation tool. This observation motivated us to design a study to explore the practices of self-declaring \aigc{} among developers. 

This study \textbf{aims} to empirically investigate the practices of self-declaring \aigc{}. Specifically, we explore the ways developers self-declare \aigc{} and the reasons behind self-declaring (or not) the use of \aigc{}. We address the following Research Questions (RQs):
\begin{itemize}
    \item \textbf{RQ1}: How do developers self-declare \aigc{} in their projects?
    \item \textbf{RQ2}: Do developers self-declare \aigc{} in practice? Why or why not?
\end{itemize} 

To address our RQs, we utilized a mixed-methods approach combining a repository mining study and a follow-up survey with developers. In the first phase of our study, we identified \aigc{} snippets from open-source projects hosted on GitHub. We collected 613 code files with \sdgc{} snippets from 586 GitHub repositories. We then analyzed and classified those code snippets. Based on the results of the mining study, we conducted a follow-up survey with practitioners to explore the ways developers self-declare \aigc{} and understand the reasons behind their choice to self-declare (or not) their \aigc{}. The survey questionnaire consists of 12 questions, including demographic questions and questions related to self-declaration practices of \aigc{}. Through the survey, we collected valid responses from 111 developers with reasonable diversity (based on location, education, experience, application domains, professional roles, and AI tool preference).   

Our \textbf{findings} show that: (1) most developers \textit{sometimes} self-declare \aigc{}, depending on the proportion of \aigc{} integrated into the project. (2) Developers use various ways to self-declare \aigc{}, ranging from \textit{simple} self-declaration comments to detailed comments that include the prompt used to generate the code, an explanation of the functionality of the code, and other code quality indications. %These ways also include self-declaring a specific code snippet or an entire code file and sharing prompts given to AI code generation tools, among others. 
(3) Developers tend to self-declare their code if the \aigc{} was not substantially modified by humans. 
%When developers use \aigc{} without substantial human modifications, they are likely to self-declare it. 
When they just use \aigc{} as a guide and then modify the code extensively, they may optionally self-declare \aigc{}. (4) As their reasons for self-declaring \aigc{}, developers reported the need for tracking and monitoring the code for future review and debugging, and also ethical considerations of using AI code generation tools. (5) Developers' reasons for not self-declaring \aigc{} include the need for extensive human modifications of \aigc{} and the perception that self-declaration is unnecessary, as they view it as comparable to seeking help from online documentation or developer forums.

The main \textbf{contributions} of this work are: (1) a curated dataset of 613 self-declared \aigc{} snippets from open-source repositories on GitHub and performing a comprehensive analysis with a focus on \sdc{} to understand the practices of self-declaring \aigc{}, (2) a practitioners’ survey to validate the findings of the mining study and explore the potential reasons behind self-declaring (or not) \aigc{}, and (3) guidelines for practitioners to self-declare \aigc{} based on the results of the mining study and the follow-up practitioners’ survey.

\textbf{Paper Organization}: Section~\ref{sec:relatedwork} discusses related work to this study. Section~\ref{sec:rqdataset} presents our research questions and outlines the methodology employed for our data collection (from the mining study and the practitioners’ survey). We present our results in Section~\ref{sec:results}, followed by the discussion of the findings and implications in  Section~\ref{sec:discussion}. We then present potential threats to the validity in Section~\ref{sec:validity} followed by the conclusion in Section~\ref{sec:conclusion}.

\section{Related Work}\label{sec:relatedwork}
Several studies have looked into different aspects of \aigc{}, in terms of the tools and model used, and their application in software engineering tasks. This includes studies that investigated the usability of AI code generation tools \cite{UsabilityOfAIProgramming,zhou2025exploring,barke2023grounded}, quality and correctness \cite{liu2024refining,ouyang2024empirical}, and security \cite{fu2023security,pearce2022asleep,majdinasab2024assessing}. Previous research has also explored the ethical aspects of using \aigc{} in software development \cite{stalnaker2024developer,xu2024distinguishing}. 

\textbf{AI Code Generation in Practice.}
One of the most studied aspects of AI code generation is its use in practice, with several surveys and user-based studies looking into adapting code generation in practice.
% Studies have investigated the usability of AI code generation in software development. 
% Vaithilingam \textit{et al.} performed a within-subject user study with 24 participants on the usability of AI code generation tools, specifically GitHub Copilot, in software development \cite{vaithilingam2022expectation}. They found that Copilot did not improve the task success rate, but developers still prefer to use it in their development because it reduces their effort to search for solutions online and provides a starting point.
Zhou \textit{et al.} investigated GitHub Copilot common problems and solutions from users' point of view, using data from GitHub issues and discussions and Stack Overflow posts \cite{zhou2025exploring}. The study identified the issues from users, including compatibility issues, copyright and policy issues, and operational issues. 
%, and their solutions are modifying configuration, using a suitable version, and reinstalling Copilot. 
Dakhel \textit{et al.} studied Copilot by generating solutions to fundamental programming problems and compared them with human solutions \cite{dakhel2023github}. They found that GitHub Copilot can produce solutions to fundamental programming problems, however, those solutions may contain bugs. Barke \textit{et al.} conducted an observation study with 20 participants while using Copilot in practice. They found that developers use these tools in two different modes: \textit{acceleration mode}, when they know how to solve the problem but use these tools to solve it quickly, and \textit{exploration mode}, where they explore various ideas to solve a problem \cite{barke2023grounded}. Similarly, Liang \textit{et al.} conducted a large-scale survey of 410 participants on the usability of AI code generation tools \cite{UsabilityOfAIProgramming}. They found that developers use these tools to enhance productivity by helping them recall syntax and reduce keystrokes. Abrahamsson \textit{et al.} studied the full-stack development capabilities of ChatGPT instead of generating small code snippets~\cite{abrahamsson2024chatgpt}. Their findings demonstrate ChatGPT's ability to create base components, decompose coding tasks, and improve development workflow, with notable improvements in ChatGPT-4 over ChatGPT-3.5, such as improved working memory and more coherent dialogues. 

\textbf{Quality, Correctness and Security of AI-Generated Code.}
Several recent studies have also investigated the quality and correctness of \aigc{}. Liu \textit{et al.} evaluated ChatGPT-generated solutions to programming problems and studied their correctness and quality (through static analysis) \cite{liu2024refining}. The results showed that ChatGPT could generate correct solutions almost 70\% of the time but suffers from maintainability issues. Ouyang \textit{et al.} studied ChatGPT code generation capabilities and found that the non-determinism of ChatGPT affects the correctness and consistency of the generated code \cite{ouyang2024empirical}. Tambon \textit{et al.} evaluated ChatGPT-generated code quality and found that ChatGPT can generate good quality code through iterative prompts, but the generated code was found to contain bugs~\cite{tambon2025bugs}, citing that ChatGPT-generated code would usually require review before integration into projects. These results show that AI-generated code will not hinder the software development process. Bui \textit{et al.} introduced a framework (called OPENIA) that uses the internal states of code-generation LLMs to evaluate the correctness of AI-generated code \cite{bui2025correctness}. This approach was found to outperform existing methods, yielding significant improvements in performance metrics and enabling more effective quality assurance in AI code generation. \textcolor{black}{Yu \textit{et al.} investigated LLM-generated code on GitHub, analyzing the characteristics of such code, its associated projects, and the nature of subsequent modifications~\cite{yu2024large}. Their findings revealed that ChatGPT and Copilot dominate LLM-generated code on GitHub, that this code is generally short and low in complexity, and that it undergoes relatively few bug-related changes.} 

Similarly, the security of the generated code is another aspect that has been widely researched. %, with studies investigating potential security limitations of AI code generation models. 
Pearce \textit{et al.} \cite{pearce2022asleep} prompted Copilot to generate code in scenarios related to high-risk security weaknesses, i.e., Common Weakness Enumerations (CWEs), covering multiple programming languages, and found that 40\% of the generated code contained security vulnerabilities. Majdinasab \textit{et al.} replicated the study by Pearce \textit{et al.} using Python examples, and found that the newer versions of Copilot improved in terms of the number of security weaknesses in the generated code. However, they still provided vulnerable code suggestions \cite{majdinasab2024assessing}. Gupta \textit{et al.} studied ChatGPT from the perspective of cybersecurity, and they found that AI tools such as ChatGPT can be used by both the defenders (secure code generation, vulnerability detection) and attackers (attack payload generation, malware code generation) \cite{gupta2023chatgpt}. Fu \textit{et al.} analyzed \aigc{} from GitHub projects to identify security weaknesses in these code snippets \cite{fu2023security}. The study shows a high likelihood of security weaknesses in \aigc{} found in public projects.

\textbf{Ethical Concerns with AI-Generated Code.}
Existing studies have explored ethical concerns with \aigc{}. A recent survey by Stalnaker \textit{et al.} with developers highlighted the ethical concerns in \aigc{}, primarily regarding intellectual property concerns related to copyright violation \cite{stalnaker2024developer}. The study reveals that developers hold mixed opinions regarding the copyright and ownership attribution of AI-generated code. 
% An important finding of their study is that developers shared the most common documentation practices, including noting the usage of AI, sharing the prompts used for code generation, and documenting the entire interaction with AI code generation tools. 
Xu \textit{et al.} investigated the ethical concerns and code quality issues with \aigc{} \cite{xu2024distinguishing}. These concerns include students' unethical use of \aigc{}, code plagiarism, and questions of code provenance. The results emphasized the importance of distinguishing \aigc{} from human-written code for code review and quality assurance processes.

\textbf{\textcolor{black}{Automatic Detection of AI-Generated Code.}}
\textcolor{black}{There has also been a body of work that explored automated approaches to detecting \aigc{}. Gurioli \textit{et al.} proposed a multilingual code stylometry approach based on transformer models that can distinguish AI-generated code from human-written code across ten programming languages with high accuracy~\cite{gurioli2025isthisyou}. Similarly, Li \textit{et al.} introduced a feature-based method for identifying ChatGPT-generated code, supported by dataset cleansing and code transformation techniques to improve detection reliability~\cite{li2023discriminating}. In contrast, our study does not focus on the automatic detection of AI-generated code but rather on developers’ self-declaration of any \aigc{} used in their projects.}

\textbf{Comparative Summary.}
Existing studies have not investigated whether developers acknowledge \aigc{}, how they acknowledge it, or what motivates them to acknowledge the use of \aigc{}. We investigate the practices of self-declaring \aigc{} among developers, examine the ways developers use to self-declare \aigc{}, including how they write the self-declaration comments and how they manage its scope, and explore the reasons behind self-declaring \aigc{}. \textcolor{black}{Existing studies, such as the work by Yu \textit{et al.}, performed a large-scale empirical analysis of GitHub repositories~\cite{yu2024large}, and the study by Stalnaker \textit{et al.}, conducted a large-scale survey and interviews with developers~\cite{stalnaker2024developer}. In contrast to the methodologies used in these existing studies, our study used a mixed-methods approach comprising GitHub repositories mining and the follow-up practitioners’ survey to gather evidence about developers' practices when self-declaring \aigc{}}. In comparison to existing work, our study addresses a unique topic using a more comprehensive research methodology. We detail our data collection and research methodology below.

\section{Research Methodology}\label{sec:rqdataset}
\begin{figure}[h]
    \centering
    \includegraphics[width=1\linewidth]{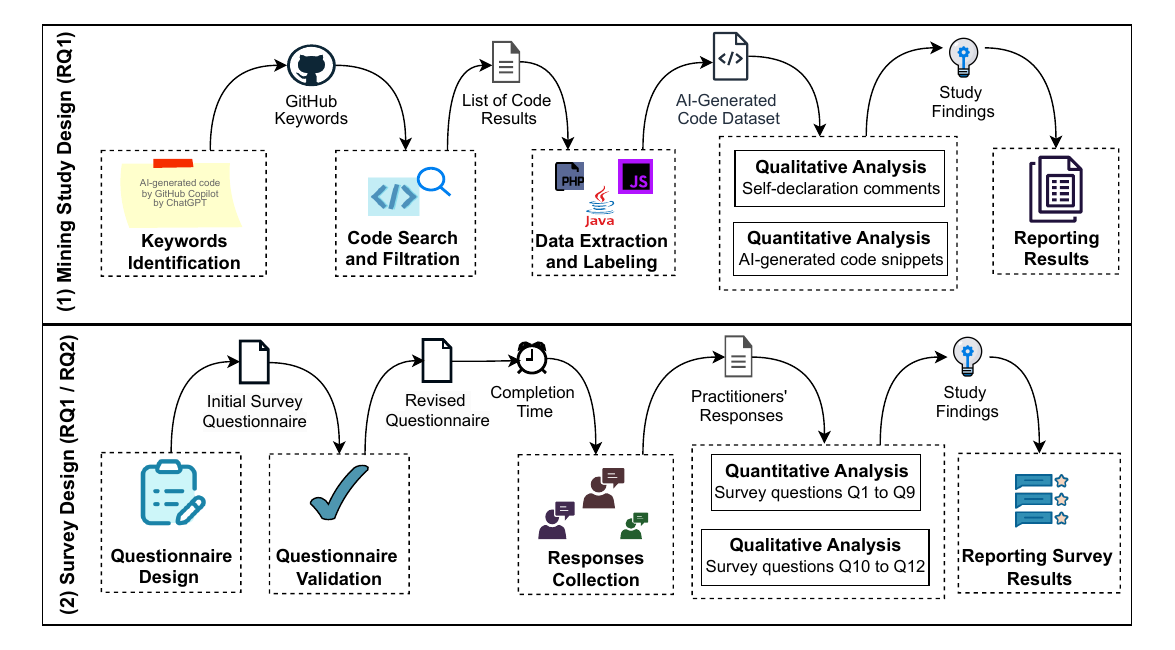}
    \caption{\textcolor{black}{Overview of the mixed-methods research process}}
    \label{fig:methodology}
    \vspace{-1em}
\end{figure}

Our study employs a mixed-methods approach comprising two main phases: (1) an analysis of \sdc{} from \aigc{} collected from open-sourced projects, and (2) a survey with practitioners regarding the practices of self-declaring \aigc{}. We first collected data about \aigc{} from real-world projects by mining GitHub repositories to answer our research question of \textit{how} \textcolor{black}{(RQ1)} developers self-declare \aigc{}. We followed that with a practitioner survey to validate the results from the mining study \textcolor{black}{(RQ1)} while addressing the research question of \textit{why} \textcolor{black}{(RQ2)} developers self-declare \aigc{}. This methodological triangulation strengthens the validity of our findings by complementing the mining study results with developers' perspectives from the practitioners’ survey. Figure~\ref{fig:methodology} shows an overview of our research methodology.

\subsection{Research Goal and Questions}\label{sec:goals}
Our study \textbf{aims} to empirically investigate \textit{the practices of self-declaring \aigc{} followed by developers}. We are interested in \textit{understanding the different ways developers self-declare \aigc{} in their projects}, and exploring \textit{the developers' reasons for self-declaring (or not) the integrated \aigc{}}. Particularly, we address the following Research Questions (RQs): 

% \subsection{Research Questions}\label{sec:researchquestions}
% \begin{tcolorbox}[arc=0mm,width=\columnwidth,
%                   top=0mm,left=0mm,  right=0mm, bottom=0mm,
%                   boxrule=.75pt]
\textbf{RQ1}: \textit{How do developers self-declare \aigc{} in their projects?}

% \end{tcolorbox}

\textbf{Motivation}. Developers may include \sdc{} in their projects to indicate AI-generated portions of code. These comments may contain important information about the generated code (for example, the tool and prompt used to generate the code). %This research question seeks to investigate the \sdc collected from GitHub repositories. We used the results of the GitHub mining study, which is the first phase of our research methodology, to answer RQ1. 
To answer this RQ, we mine GitHub repositories to locate code comments that may indicate if the code is AI-generated. We refer to those comments as \textit{\sdc{}}. \textcolor{black}{We also confirmed the findings we obtained from the mining study (RQ1) with a follow-up survey with practitioners.} RQ1 is further divided into five sub-RQs. 

\textbf{RQ1.1}: \textit{What information do developers share in their \sdc{}?}

This RQ focuses on studying the content of the \sdc{} collected from the projects. Developers may share useful information about \aigc{} in the added code comments (e.g., the AI tool/model used, the prompt used to generate the code, additional code explanations, and code quality indications). The information shared in these comments may provide insights into how developers write \sdc{}. 

\textbf{RQ1.2}: \textit{What is the scope of the \sdc{}?}

This RQ focuses on analyzing the scope of the \sdc{} based on information such as where they are mentioned (i.e., the position of the comment) in the file. \textcolor{black}{The scope of the self-declaration comments impacts the traceability and maintainability of the \aigc{}. A broad scope (file-level) may compromise accurate traceability, while a granular scope (snippet-level) may lead to redundancy and inefficiency. By analyzing common developer practices, this RQ provides guidelines for effective scope management in self-declaring \aigc{}.}

% The scope of self-declaration comments impacts the traceability and maintainability of AI-generated code. Broad scope (file-level) of self-declaration comments may risk accurate traceability, while granular scope (snippet-level) of self-declaration comments may cause redundancy and inefficiency. By analyzing the common ways of developers, this RQ informs guidelines for effective scope management in self-declaring AI-generated code.

% The answer to this RQ helps in understanding the common ways developers use to manage the scope of the \sdc{}, which is valuable for accurately identifying \aigc{}.

\textbf{RQ1.3}: \textit{What is the distribution of the \aigc{} snippets with \sdc{} across repositories?}

This RQ focuses on analyzing two key aspects: (1) the distribution of the \sdc{} within individual source files and (2) the number of files containing \sdc{} within a single repository. This information helps quantify the presence of self-declared \aigc{} within repositories.
% , offering meaningful information on its adoption by developers in practice.
\textcolor{black}{The presence and frequency of self-declaration comments in GitHub repositories can serve as indicators of how widely developers have adopted the self-declaration practices for AI-generated code. This RQ aims to illustrate the trend in adopting these self-declaration practices over the years.}
% The presence and frequency of self-declaration comments in GitHub repositories can serve as indicators of how widely developers have adopted the practices of self-declaring \aigc{}. This RQ also aims to illustrate the trend in adopting these self-declaration practices over the years.

\textbf{RQ1.4}: \textit{Are developers satisfied with the \aigc{} according to \sdc{}?} 

This RQ aims to investigate developers' sentiments by categorizing \sdc{} that contain sentiment information as either positive or negative. A positive self-declaration reflects developers' satisfaction with the integrated \aigc{}, whereas a negative self-declaration indicates developers' dissatisfaction with the integrated \aigc{}. %The answer to this RQ provides insights into developers' satisfaction regarding the \aigc{} they have integrated into their projects.

% \begin{tcolorbox}[arc=0mm,width=\columnwidth,
%                   top=0mm,left=0mm, right=0mm, bottom=0mm,
%                   boxrule=.75pt]

% \textbf{RQ2}: \textit{Do developers self-declare \aigc{} in practice? Why or why not?}
% \textbf{RQ2}: \textit{Do developers frequently self-declare \aigc{} in practice? Why or why not?}
\textcolor{black}{\textbf{RQ2}: \textit{What proportion of developers self-declare \aigc{} in practice? Why or why not?}}

% \end{tcolorbox}

\textbf{Motivation}. 
AI code generation tools can speed up development and boost developers' productivity \cite{generativeAIforPractioners,UsabilityOfAIProgramming}. Developers increasingly use AI code generation tools in practice \cite{SO-survey2024}. 
% Here, we aim to investigate whether developers self-declare any \aigc{} used in their projects.
\textcolor{black}{Therefore, we aim to investigate what proportion of developers self-declare any \aigc{} used in their projects and how frequently they self-declare such code.}
We also explore the underlying reasons for developers to self-declare and not to self-declare \aigc{}. The answer to this RQ helps understand the current situation of self-declaration of \aigc{} as a practice in software development. RQ2 is further divided into the following two sub-RQs:

\textbf{RQ2.1}: \textit{How common do developers self-declare \aigc{} in their projects?}

This RQ investigates how widely \textcolor{black}{(the proportion of)} developers self-declare (or not) \aigc{} in their projects \textcolor{black}{from our practitioners’ survey}. \textcolor{black}{The answer to this RQ aims to provide insights into the current state of self-declaration practices of AI-generated code among developers.}
% This RQ helps us understand the distribution of developers who self-declare versus those who do not self-declare the use of AI-generated code in their projects.

\textbf{RQ2.2}: \textit{What are the reasons developers self-declare (or not) the \aigc{} in their projects?}

This RQ explores the underlying reasons why developers self-declare (or not) \aigc{} in their projects. \textcolor{black}{This RQ aims to identify the potential motivation of developers to self-declare \aigc{} and any barriers that lead to the decision of not self-declaring \aigc{}.}

\subsection{Mining Study Design} \label{subsec:Miningstudydesign}
To answer RQ1, we collected self-declared \aigc{} from GitHub repositories. We defined a \textit{self-declared \aigc{}} snippet as one in which the developer explicitly acknowledges that the code was written with the assistance of an AI code generation tool. This self-declaration is expressed in the form of code comments added anywhere in the file. We collected self-declared \aigc{} from GitHub because it is the most widely used platform by developers to host code \cite{Octoverse2024}. This allowed us to access code repositories and study projects written in various languages.

\textcolor{black}{We used a keyword-based search to locate code snippets generated by four well-known AI code generation tools: Copilot, ChatGPT, CodeWhisperer, and Code Llama. We prioritized Copilot and ChatGPT due to their popularity and widespread use by developers in practice \cite{UsabilityOfAIProgramming, SO-survey2024, yu2024large}}. %In addition to these two tools, we also included search terms for CodeWhisperer and Code Llama to ensure a diverse dataset of \aigc{} in this study. 

\subsubsection{\textcolor{black}{Keyword Identification}} 
\label{subsubsec:keyword}
The first step of the data collection process is to identify the search terms to retrieve \aigc{} from GitHub projects. For this purpose, we performed a pilot data search to determine the most suitable search terms. We used the names of popular AI code generation tools to locate \aigc{} snippets. However, using only the tool names mainly returned the projects where developers shared their experience using AI code generation tools, and did not yield code snippets generated by these tools. We also examined the use of more generic search terms without specifying the name of the tools used (e.g., ``\textit{\aigc{}}'', ``\textit{AI-assisted code}''), which returned only a few \aigc{} snippets. We finally noted that using specific search terms that include the name of the tools (e.g., ``\textit{generated by GitHub Copilot}'', ``\textit{written by ChatGPT}'') can return more relevant results than using those generic search terms. The search terms appreciating AI code generation tools, such as ``\textit{thanks Copilot}'' and ``\textit{thanks ChatGPT}'' also returned relevant results. \textcolor{black}{We also searched using \textit{warning} search terms, such as ``\textit{warning ChatGPT}'' and ``\textit{caution Copilot}''. However, we were not able to find a considerable number of results using those search terms.} \textcolor{black}{The four search strategies and the pilot search terms are listed in Table \ref{tab:pilot_search_terms}.}

\begin{table}[htbp]
\centering
\small
\begin{tabular}{@{}p{0.03\linewidth}p{0.22\linewidth}p{0.7\linewidth}@{}}
\toprule
\textbf{\#} & \textbf{Search Strategy} & \textbf{Example Terms} \\ \midrule
1 & Tool Names Only & 
ChatGPT, GitHub Copilot, Code Llama, Amazon CodeWhisperer \\

2 & Generic AI Terms & 
AI-generated code, AI-assisted code, code generated by AI, code written with the help of AI \\

3 & Tool-Specific Patterns & 
generated by \{tool name\}, written by \{tool name\}, from \{tool name\}, by \{tool name\}, solution by \{tool name\}, help of \{tool name\}, coded by \{tool name\} \\

4 & Warning and Thanking Terms & 
thanks \{tool name\}, credit to \{tool name\}, caution \{tool name\}, warning \{tool name\} \\
\bottomrule
\end{tabular}
\caption{\textcolor{black}{Search strategies and initial pilot search terms to identify the search terms for data collection}}
\label{tab:pilot_search_terms}
\vspace{-2em}
\end{table}

We performed the pilot search using GitHub's REST API \cite{codeSearchAPI}. As a result, we identified \aigc{} at the repository and code label. We classified the results as \textit{Repository label} when the repositories explicitly acknowledge using \aigc{} in their project descriptions (e.g., in the \texttt{README} files). On the other hand, we classified results as \textit{Code label} when the code files and snippets contain \sdc{} about the \aigc{}. The first author classified each code snippet (e.g., whether it is AI-generated based on self-declaration comments), and then the results were cross-validated with two other co-authors. Through this process, we noticed that the search results from the repository label were more likely to contain dummy and student practice projects (i.e., small experimental projects with little content). Considering this, we decided to limit our data collection to the search results from the code label to ensure the quality of our dataset.

In the search results from the code label, we noticed that developers included \sdc{} in different positions of the code files, acknowledging that the code is AI-generated. The first author collected the code snippets from search results that mention the self-declaration of \aigc{}.

\textcolor{black}{In the pilot search phase, we collected a significant number of self-declared AI-generated code using search terms that included the names of the AI code generation tools and phrases specific to code generation. Therefore, out of the total 23 search terms finally selected for our data collection, 20 were based on Strategy 3 (see Table 1). The list included only 2 ``\textit{thanking}'' terms from Strategy 4 because we also found relevant results using these two terms during the pilot phase. We only included one general search term without the AI tool name (Strategy 2) because we found limited results for such search terms. We could not find any results using the terms from Strategy 1, and we did not include any search terms from this strategy. Based on the results of the pilot search, 22 search terms included the names of the AI code generation tools. 17 out of the 22 search terms included names of \textit{ChatGPT} and \textit{GitHub Copilot}, and only 5 search terms included names of other tools.} The list of search terms is provided in Table~\ref{tbl:Keywords}, which also shows the number of \aigc{} snippets included in our dataset. 

% \begin{table}[]
% \small
% \footnotesize
% \setlength{\tabcolsep}{10pt}
% \renewcommand{\arraystretch}{1.3} 
% \begin{tabular}{|>{\raggedright\arraybackslash}p{0.05\textwidth}|>{\raggedright\arraybackslash}p{0.55\textwidth}|>{\raggedright\arraybackslash}p{0.2\textwidth}|}
% \hline
% \textbf{\#} & \textbf{Search Terms} & \textbf{Included Results} \\ \hline
% 1 & written by ChatGPT & 132 \\ \hline
% 2 & generated by ChatGPT & 111 \\ \hline
% 3 & written by Copilot & 86 \\ \hline
% 4 & thanks ChatGPT & 66 \\ \hline
% 5 & generated by GitHub Copilot & 66 \\ \hline
% 6 & thanks Copilot & 41 \\ \hline
% 7 & by GitHub Copilot & 21 \\ \hline
% 8 & AI-generated code & 19 \\ \hline
% 9 & solution by ChatGPT & 13 \\ \hline
% 10 & from ChatGPT & 11 \\ \hline
% 11 & help of GitHub Copilot & 10 \\ \hline
% 12 & solved by GitHub Copilot & 6 \\ \hline
% 13 & coded by GitHub Copilot & 6 \\ \hline
% 14 & solution by GitHub Copilot & 4 \\ \hline
% 15 & from GitHub Copilot & 4 \\ \hline
% 16 & generated by CodeWhisperer & 4 \\ \hline
% 17 & from AWS CodeWhisperer & 1 \\ \hline
% 18 & created by GitHub Copilot & 1 \\ \hline
% 19 & using GitHub Copilot & 1 \\ \hline
% 20 & by CodeWhisperer & 0 \\ \hline
% 21 & by Llama & 0 \\ \hline
% 22 & from Code Llama & 0 \\ \hline
% 23 & proposed by GitHub Copilot & 0 \\ \hline
% \end{tabular}
% \caption{Search terms used for searching AI-generated code from GitHub}
% \label{tbl:Keywords}
% \end{table}

% \usepackage{booktabs}
% \usepackage{graphicx}
\begin{table}[h]
% \small

% \footnotesize
% \setlength{\tabcolsep}{10pt}
% \renewcommand{\arraystretch}{1.3}
\adjustbox{max width=\textwidth}{%
\small
\begin{tabular}{@{}lll|lll@{}}
\toprule
\textbf{\#} & \textbf{Search Term} & \textbf{Results} & \textbf{\#} & \textbf{Search Term} & \textbf{Results} \\ \midrule
1 & written by ChatGPT & 132 & 13 & coded by GitHub Copilot & 6 \\
2 & generated by ChatGPT & 111 & 14 & solution by GitHub Copilot & 4 \\
3 & written by Copilot & 86 & 15 & from GitHub Copilot & 4 \\
4 & thanks ChatGPT & 66 & 16 & generated by CodeWhisperer & 4 \\
5 & generated by GitHub Copilot & 66 & 17 & from AWS CodeWhisperer & 1 \\
6 & thanks Copilot & 41 & 18 & created by GitHub Copilot & 1 \\
7 & by GitHub Copilot & 21 & 19 & using GitHub Copilot & 1 \\
8 & AI-generated code & 19 & 20 & by CodeWhisperer & 0 \\
9 & solution by ChatGPT & 13 & 21 & by Llama & 0 \\
10 & from ChatGPT & 11 & 22 & from Code Llama & 0 \\
11 & help of GitHub Copilot & 10 & 23 & proposed by GitHub Copilot & 0 \\
12 & solved by GitHub Copilot & 6 & & & \\ 
\bottomrule
\end{tabular}%
}
\caption{\textcolor{black}{Search terms used for identifying AI-generated code in GitHub repositories}}
\label{tbl:Keywords}
\vspace{-2em}
\end{table}

\subsubsection{Search and Filtration Process}
We applied the finalized search terms to locate relevant code snippets from GitHub projects. The search results were filtered based on the programming languages used in the projects. To increase the language diversity in our dataset, we collected code files from the top 10 programming languages on GitHub ~\cite{TopLanguages}: Python, Java, JavaScript, TypeScript, C, C++, Go, PHP, Ruby, and C\#. We used the search terms we identified in Table~\ref{tbl:Keywords} and applied GitHub’s programming language filter for the top 10 languages. In most cases (19 search terms), the search returned valid results. \textcolor{black}{There were four search terms for which we were unable to retrieve any results that met our inclusion criteria (see Table \ref{tbl:Keywords}).}

We manually checked the source code files for \sdc{} to determine whether they contain \aigc{}. During this process, we applied our inclusion and exclusion criteria (see Table \ref{tab:miningcriteria}). First, we only included code snippets in our dataset when it is clearly stated in the \sdc{} that they are AI-generated. Secondly, we only included \aigc{} snippets that belong to a repository with more than 1,000 lines of code to avoid dummy and small student practice projects (we used \texttt{cloc}\footnote{\url{https://github.com/AlDanial/cloc.git}} to calculate the lines of code in each repository). 

\begin{figure}[h]
    \centering
    \includegraphics[width=0.8 \linewidth]{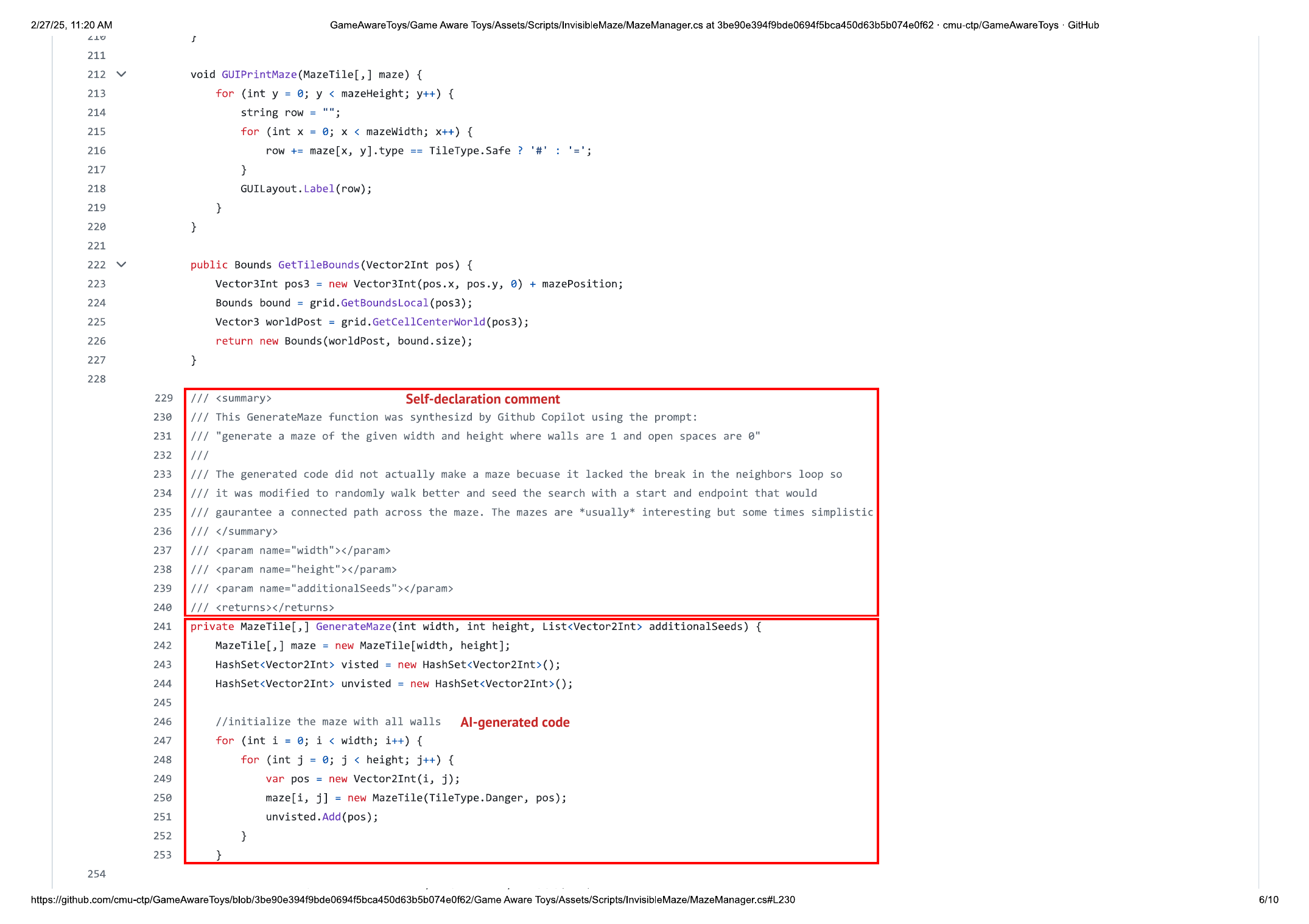}
    \caption{\textcolor{black}{Example of included \aigc{} demonstrating inclusion criterion I1}}
    \label{fig:Inclusion}
\end{figure}

The inclusion or exclusion criteria are provided in Table \ref{tab:miningcriteria}. 
Figure~\ref{fig:Inclusion} shows an example of \aigc{} that meets our inclusion criterion I1 because the developer declared in the self-declaration comment that the \texttt{GenerateMaze()} function is generated by GitHub Copilot. It also meets our inclusion criterion I2 because the code is part of the repository \texttt{GameAwareToys} that contains 98 files with 46,198 lines of code, exceeding the 1,000 lines of code threshold.

% \textcolor{red}{Figure \ref{fig:Exclusion} shows an example of the exclusion criterion E1 (see Table \ref{tab:miningcriteria}), as the self-declaration comment did not reveal if the code is AI-generated, but only contained the word ``copilot''. Additionally, it belongs to the repository \texttt{elysia-lambda}, which has only 456 lines of code, which is less than the 1000 lines of code threshold (exclusion criterion 2).}

% \begin{figure}[h]
%     \centering
%     \includegraphics[width=0.8 \linewidth]{Figures/exclusion-criteria.pdf}
%     \caption{Example of excluded \aigc{} demonstrating exclusion criterion E1}
%     \label{fig:Exclusion}
%     \vspace{-1em}
% \end{figure}

Once the \aigc{} is identified, we manually reviewed the repositories associated with these code files. To collect a high-quality dataset, we gathered demographic information about the repositories where the \aigc{} was retrieved, including the \textit{number of commits}, \textit{number of contributors}, and \textit{application domain}. \textcolor{black}{We collected this information to illustrate the maturity and activity level of the repositories. Although this information is not directly tied to our RQs, it provides useful context for understanding the results and characteristics of our dataset, showing the relevance and representativeness of mined repositories.} We also collected the links to the source code files and \sdc{}.

\begin{table}[htbp]
\footnotesize
\setlength{\tabcolsep}{3pt} 
\renewcommand{\arraystretch}{0.9} 
\begin{tabular}{>{\raggedright\arraybackslash}p{0.05\linewidth}>{\raggedright\arraybackslash}p{0.85\linewidth}}
\toprule
\multicolumn{2}{l}{\textbf{Inclusion Criteria}} \\
\midrule
I1 & The self-declaration comment clearly mentions the code as AI-generated. \\
I2 & The source code belongs to a project with more than 1,000 lines of code. \\
\midrule
\multicolumn{2}{l}{\textbf{Exclusion Criteria}} \\
\midrule
% E1 & The self-declaration comment does not clearly indicate that the code is AI-generated. \\
E1 & The code belongs to a project likely to be a dummy or student project with less than 1,000 lines of code. \\
\bottomrule
\end{tabular}
\caption{Inclusion and exclusion criteria for mining \aigc{} from GitHub repositories}
\label{tab:miningcriteria}
\vspace{-2em}
\end{table}

We categorized the search results into two main categories: \snl{} and \fl{} self-declarations. Snippet-level is when the self-declaration comment is placed \textit{before} the code snippet, while \fl{} is when the self-declaration comment is placed at the \textit{beginning} or \textit{end} of the source code file. We also labeled self-declaration comments in which developers' sentiment regarding the \aigc{} is evident in the \sdc{}. We manually categorized the comments in which developers appreciated the assistance of AI code generation tools as \textit{positive} and the comments in which developers shared warnings and disclaimers about the use of \aigc{} as \textit{negative}. All of this information was recorded in an MS Excel file, which is included in our dataset~\cite{replicationPackage}.

During the data labeling process, the first author performed the labeling, and then the other co-authors cross-validated the labeling results. \textcolor{black}{We calculated the agreement level between the authors using the average pairwise Cohen’s Kappa coefficient \cite{cohen1960cofficient}, by computing Cohen’s Kappa for each pair of the three coders and then reporting the average value. The resulting Cohen’s Kappa coefficient was 0.82, indicating a high level of agreement among the authors.}

\textcolor{black}{We initially curated slightly over 1,200 code files from public GitHub repositories. After applying our inclusion and exclusion criteria (see Table \ref{tab:miningcriteria}), we narrowed down the selection to 899 files that fulfilled our criteria. We retained 613 code files from 586 unique repositories using manual sub-sampling to ensure a diverse dataset for data analysis. We included a few representative files from repositories containing multiple files with \aigc{}.} This was necessary as several repositories contained many files with highly similar self-declaration comments. For instance, the repository \texttt{KV6002-NUMUNCHIES} alone contained 65 files with similar self-declaration comments with small variations. On the contrary, for the analysis of the distribution of AI-generated code in RQ1.3, we recorded the total number of files containing self-declared AI-generated code snippets in the ``\textit{Total files with SDC}'' column (\texttt{code-details.xlsx}) in our dataset \cite{replicationPackage}.

\subsubsection{Data Extraction}
% In the data extraction, we focused on extracting important information about \aigc{}, as illustrated in Table. 
 %Since the goal of our study is to explore self-declaration of \aigc{}, we collected \sdc{} mentioned by developers in the code files.
%These comments were recorded in the \textit{code-details} spreadsheet file without any changes. We recorded the background of this data collection, such as the search term and the programming language filter used to find the code result. We recorded if the comments were \snl{} or \fl{} to gather quantitative data about the code. 
During data extraction, we first reviewed each repository's description (\texttt{README} file) to understand the nature of the projects and their application domains. We also extracted other demographic information, including the number of contributors, the number of commits, and the AI code generation tools used. %To make the source code files easily accessible, 
Secondly, we extracted the repository name, programming languages, the line numbers, the number of self-declaration comment instances in a single file, and the total number of files containing \aigc{} snippets in a repository. We ensured that we documented the exact line in the code file where the self-declaration comment was mentioned. \textcolor{black}{We calculated the Lines of Code (LoC) for all 613 code files by retrieving their content using the GitHub REST API~\cite{codeSearchAPI} and counting the number of lines in each file.} \textcolor{black}{We also extracted the dates when lines containing self-declaration comments were added to code files in our dataset to analyze trends of self-declaring AI-generated code over time. This was achieved using the GitHub blame object obtained through the GitHub GraphQL API~\cite{GraphQLAPI}. In total, we examined 613 files and successfully collected timestamps for 601 files. The remaining 12 files were excluded because their repositories had been made private.} This recorded information was constantly verified and reviewed by all co-authors. Table~\ref{tbl:Extraction} provides an overview of our data extraction.

\begin{table}[htbp]

\resizebox{\columnwidth}{!}{%
\begin{tabular}{@{}llll@{}}
\toprule
\textbf{\#} & \textbf{Data Item} & \textbf{Description} & \textbf{Related RQ} \\ \midrule
D1 & Self-declaration comments & Self-declaration comments that indicate the code as AI generated. & RQ1.1, RQ1.4 \\
D2 & Repository name & The name of the repository to which the code snippet belongs. & General \\
D3 & File name & File name and direct link to the source code file. & General \\
D4 & Application domain & Description and the project. & Demographic \\
D5 & Programming Language & Main programming language of code snippet. & Demographic \\
D6 & Code commits and contributors & Statistics obtained from the repositories. & Demographic \\
D7 & Keyword used & Keywords used to retrieve the code snippets. & General \\
D8 & Total files with SDC & The total files containing \sdc{}. & RQ1.3 \\
D9 & Multiple SDC in a single file & Extracted multiple \sdc{} in a single source code file. & RQ1.3 \\
D10 & Comment type & Classification of the comment (snippet-level or file-level). & RQ1.2 \\
D11 & Commit date & The date when self-declaration comment was added to the code. & RQ1.3 \\ 
D12 & LoC  & The number of lines of code in a code file. & RQ1.2 \\
\bottomrule
\end{tabular}%
}
\caption{\textcolor{black}{Data items extracted from GitHub repositories}}
\label{tbl:Extraction}
\vspace{-2em}
\end{table}

\subsubsection{Data Analysis}
% We employed grounded theory to qualitatively analyze our data~\cite{stol2016grounded}. Grounded theory is a bottom-up approach emphasizing theory generation instead of extending or verifying existing theories~\cite{stol2016grounded}. Specifically, we employed open coding and constant comparison~\cite{stol2016grounded} to analyze \sdc{} data collected from GitHub repositories. 
\textcolor{black}{We employed thematic analysis as our main methodology. We adopted a predominantly inductive and data-driven approach to provide a comprehensive interpretation of the data~\cite{braun2006using}. This method allowed us to systematically identify, analyze, and organize patterns in the \sdc{} collected from GitHub repositories.} 

% changed open coding to initial coding as open coding is more associated with GTM
\textcolor{black}{\textit{Initial Coding.} We performed the initial coding by identifying and labeling essential ideas and concepts from \sdc{}. The first author carefully read each self-declaration comment to extract codes. We iteratively refined the code as we examined more data. The other co-authors continuously reviewed the codes and shared their feedback to improve the reliability of the coding process. \textcolor{black}{The first author shared the initial coding results with the co-authors through a collaborative document. The refinement process occurred in two rounds. First, the co-authors reviewed the document and provided written comments. Then, any remaining disagreements were discussed and resolved during live meetings.} \textcolor{black}{This approach aligns with established qualitative research practices, where a single primary coder conducts the initial coding while co-authors engage in iterative review and discussion to enhance the reliability of the coding process \cite{McDonald2019reliability}.} The analysis was done using MAXQDA\footnote{\url{https://www.maxqda.com/}} to organize and manage the coding results.}

\textcolor{black}{As an example of initial coding, we labeled the self-declaration \href{https://github.com/Kashifraz/SDAGC/blob/main/Mining\%20Study/python/decompositions.py\#L3173}{comment}, ``\textit{NB: Much of this was written with Copilot! (although still had to fix a bunch of issues)}'' as a \textit{negative code quality indication}. Similarly, we labeled the following self-declaration \href{https://github.com/Kashifraz/SDAGC/blob/main/Mining\%20Study/javascript/bitcoinController.js\#2}{comment}, ``\textit{create a public function that handles request to the Bitcoin API, this code is generated by GitHub Copilot}'' as a \textit{short code explanation}.} 

% added wording "themes" instead of "concepts and categories"
\textcolor{black}{\textit{Iterative Themes Development.} We developed high-level themes by iteratively and constantly comparing the codes in the initial coding phase. 
% These two sentences can be dropped as we have changed from GTM to thematic analysis.
% This was performed iteratively as we compared the codes in individual \sdc{} (comments under study) with the codes from all \sdc{} we have encoded so far. The purpose of comparing codes against each other is to group codes with similar meanings into high-level themes.
We developed a total of 7 high-level themes from the qualitative analysis of \sdc{}. For instance, we categorized all positive and negative code quality indications mentioned in \sdc{} into the \textit{Code Quality Indications} theme. Similarly, we categorized the \sdc{} containing various code descriptions and explanations into the \textit{Code Explanations} theme. The first author did the initial coding, while the other two co-authors provided feedback on these themes by reviewing the coding. We resolved disagreements using the negotiated agreement approach \cite{campbell2013coding} to enhance the reliability of our data analysis.}

\subsection{Survey Design} \label{subsec:surveydesign}
\begin{figure}[h]
    \centering
    \includegraphics[width=0.8\linewidth]{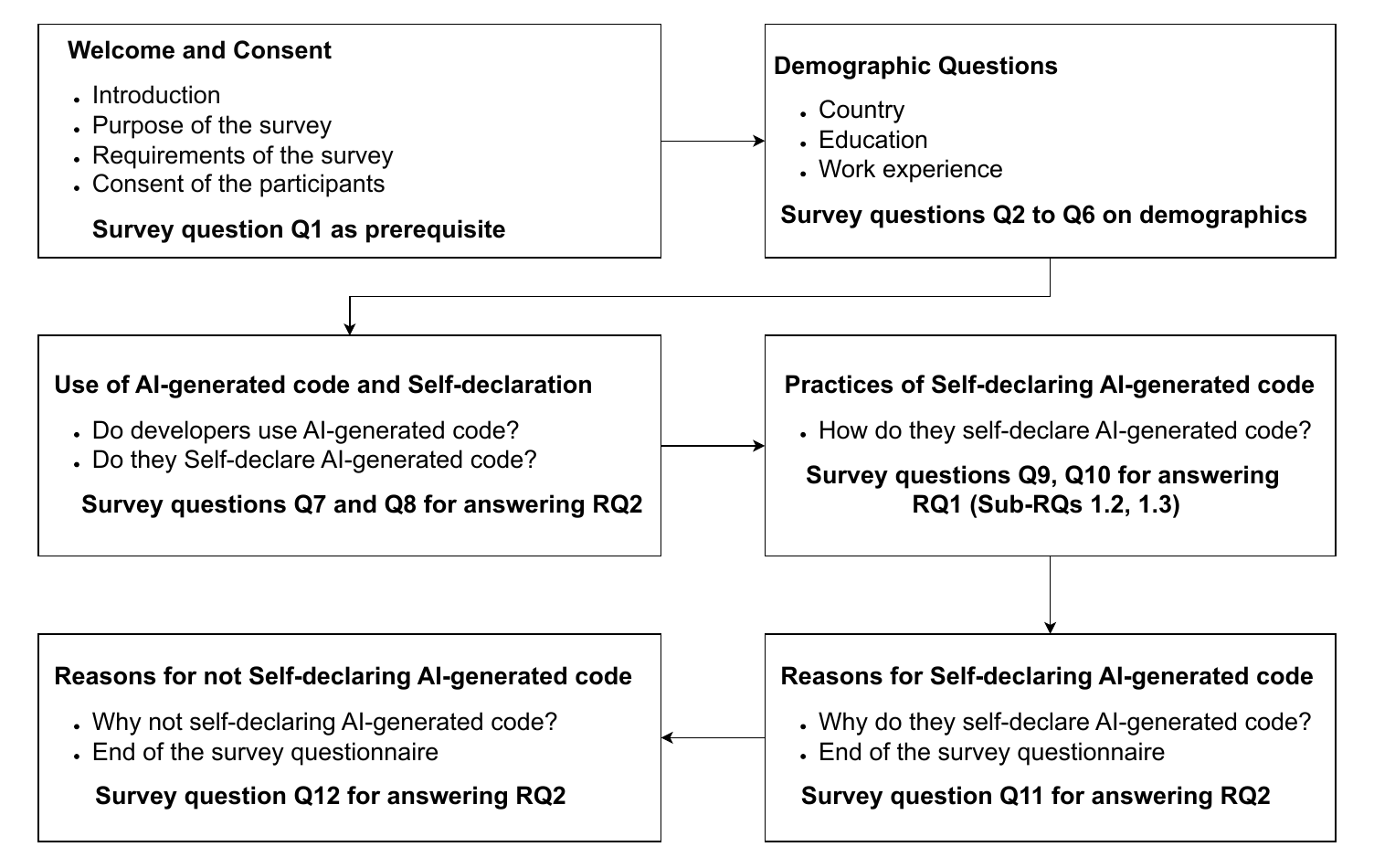}
    \caption{Overview of the survey questionnaire}
    \label{fig:SurveyContent}
    \vspace{-1em}
\end{figure}

The second phase of our study is the practitioner survey, where we investigate how and why developers self-declare their \aigc{}. The survey is motivated by the findings of the mining study, described above in Section \ref{subsec:Miningstudydesign}. %The detailed survey design is shared below. 
%\subsubsection{Data Collection}
%The survey method is used to collect data from practitioners. 
The survey workflow is shown in Figure~\ref{fig:SurveyContent}. We conducted the survey online to allow easy distribution to a vast number of participants. Our questionnaire includes both open-ended and closed-ended questions. %This enabled us to acquire both quantitative and qualitative information to answer our research questions.

\subsubsection{Questionnaire Design} 
We designed the questionnaire with a clear focus on our research objectives to ensure that each Survey Question (SQ) is directly aligned with our Research Questions (RQs). The survey begins with an information page that explains the study's context and outlines the requirements for participating in this survey. The questionnaire is structured into five sections, as follows:  
\begin{enumerate}
    \item \textit{Demographic questions}: General information about the participants, including educational background, current position, location, and experience.
    \item \textit{Experience with AI code generation tools}: Reflection on developers' experience and preferences regarding the use of AI code generation tools in practice. 
    \item \textit{Ways of self-declaring \aigc{}}: Exploration of the various ways developers use to self-declare their \aigc{}.
    \item \textit{Reasons for self-declaring \aigc{}}: Examination of the motivations behind developers' decisions to self-declare their \aigc{}.
    \item \textit{Reasons for not self-declaring \aigc{}}: Examination of the motivations behind developers' decisions not to self-declare \aigc{}.
\end{enumerate}

We provide an overview of our survey questionnaire in Table~\ref{tab:survey}, which is also accessible online\footnote{\url{https://forms.gle/WUaFEKBPpfndLfXf8}}. \textcolor{black}{We distributed the survey through our connections and various mailing lists, specifically targeting software developers who have experience using AI code generation tools to ensure a relevant study population. The survey was open for participants until mid-October 2024.}

\begin{table}[htbp]

\resizebox{\columnwidth}{!}{%
\begin{tabular}{@{}llll@{}}
\toprule
\textbf{ID} & \textbf{Section} & \textbf{Question} & \textbf{Data Analysis Method} \\ \midrule
SQ1 & Introduction & Do you use AI-generated code in your projects? & Descriptive Statistics \\
SQ2 & Demographic & Which country are you working in? & Descriptive Statistics \\
SQ2 & Demographic & What is your highest educational degree? & Descriptive Statistics \\
SQ4 & Demographic & How many years of experience do you have? & Descriptive Statistics \\
SQ5 & Demographic & What is your role in your organization? & Descriptive Statistics \\
SQ6 & Demographic & \begin{tabular}[c]{@{}l@{}}What are the domains or areas of your software projects \\ in which AI-generated code is declared?\end{tabular} & Descriptive Statistics \\
SQ7 & RQ1, RQ2 & Which AI-code generation tools you use in your projects? & Descriptive Statistics \\
SQ8 & RQ2 & \begin{tabular}[c]{@{}l@{}}When you use an AI-code generation tool to generate code, \\ do you self-declare it in your projects?\end{tabular} & Descriptive Statistics \\
SQ9 & RQ1 & \begin{tabular}[c]{@{}l@{}}If you self-declare AI-generated code, how do you self-declare \\ them in your projects?\end{tabular} & Descriptive Statistics \\
SQ10 & RQ1 & \begin{tabular}[c]{@{}l@{}}Why do you use the practice(s) you mentioned in previous \\ question to self-declare AI-generated code?\end{tabular} & Open Coding and Constant Comparison \\
SQ11 & RQ2 & \begin{tabular}[c]{@{}l@{}}What are the reasons for self-declaring AI-generated code \\ in your project?\end{tabular} & Open Coding and Constant Comparison \\
SQ12 & RQ2 & What are the reasons for not self-declaring AI-generated code? & Open Coding and Constant Comparison \\ \bottomrule
\end{tabular}%
}
\caption{An overview of survey questions and their analysis method}
\label{tab:survey}
\vspace{-2em}
\end{table}

We followed a thorough review and validation process to ensure the survey questionnaire captured the necessary information. %First, the initial questionnaire was created by the first author and then reviewed by the other two co-authors. 
%These reviews focused on improving the readability and comprehensibility of the survey questionnaire. 
We carefully reviewed the questionnaire's content and structure, and cross-validated its contents to enhance clarity and relevance.
We then sought the expertise of an external researcher (non-author) to validate the clarity and rationale of the survey. To that end, we invited a senior researcher with strong expertise in qualitative software engineering research to review the questionnaire for enhancing its clarity. Following his comments, we revised the structure and questions to improve the survey clarity. For instance, we added a prerequisite question to ask participants if they actively use AI code generation tools in practice. We also decomposed the survey into smaller sections to improve readability. %For example, we created separate sections for the questions about reasons for self-declaring \aigc{} and reasons for not self-declaring \aigc{}.

Next, we conducted a pilot run of the survey by distributing the questionnaire to a small subset of our target population. These participants were primarily practitioners with whom we had direct contact in the industry, allowing us to easily access them for feedback and suggestions. We received 11 valid responses from the participants of the pilot survey and made several revisions to the survey questionnaire to improve its structure and readability. %We shared on the Welcome page of the questionnaire that our survey consists of a total of 12 questions organized into five sections, and participants may need about 5 minutes to complete it. We also mentioned that our survey invites participants who are professionally engaged in software development and utilize \aigc{} in their projects. We proceeded with the formal data collection with these enhancements to the survey questionnaire.

\subsubsection{Data Collection} 
We identified our target population as software engineers who potentially use AI code generation tools in their practice. We distributed the survey to the target population, seeking responses from diverse groups with different experience levels and roles in their organizations. We did not impose any restrictions on participants based on their job titles, geographical location, educational qualifications, or levels of experience. 

% This approach facilitated the collection of diverse responses, enriching the overall findings of our research.

%\noindent\textbf{Sampling and Population.} 
One challenge we encountered was the difficulty in employing a probabilistic sampling method. It is not feasible to contact every individual within our target population for participation in the survey. As a result, we opted for non-probabilistic sampling methods, which include convenience and snowball sampling. We implemented four main strategies to effectively reach our target population and ensure that our sample accurately reflects the overall population: 

\begin{enumerate}
    \item Recruit practitioners from our contacts who are willing to contribute to our survey, utilizing the convenience sampling method.
    \item Invite contributors of the repositories we collected during our mining study to participate in our survey using purposive sampling.
    \item Request participants to invite practitioners to participate in our survey using a snowball sampling approach.
    \item Invite participants through social media platforms such as Facebook and LinkedIn using convenience sampling.
\end{enumerate}

These three strategies were designed to facilitate our data collection process while acknowledging the limitations inherent in non-probabilistic approaches. %We followed the strategies shared above to recruit participants for our survey. 
We sent invitation emails to the contributors of the repositories we collected during our mining study. We invited practitioners in our network via email and LinkedIn messages. We also invited participants through survey invitations posted on our social media profiles and various developer communities and groups on social media platforms, such as LinkedIn and Facebook. \textcolor{black}{Our survey received a total of 81 responses from social media (primarily LinkedIn, with some from Facebook), 36 responses from participants contacted directly, and 21 responses from contributors to the repositories of our mining study dataset. Participation was entirely voluntary, and no compensation was provided.}

\subsubsection{Data Analysis} \label{subsection:surveydataanalysis} 
% We explain our data analysis process for the survey responses in detail below.  

\textit{Validating Responses.} To maintain the reliability of our data, we developed a set of inclusion and exclusion criteria to review and validate each survey response (as shown in Table \ref{tab:surveycriteria}). \textcolor{black}{Although we did not explicitly screen for AI-generated spam responses, the manual review process substantially reduced the likelihood of low-quality or automated submissions.}  
 
\begin{table}[htbp]
\footnotesize
\setlength{\tabcolsep}{3pt} 
\renewcommand{\arraystretch}{0.9} 
\begin{tabular}{>{\raggedright\arraybackslash}p{0.05\linewidth}>{\raggedright\arraybackslash}p{0.85\linewidth}}
\toprule
\multicolumn{2}{l}{\textbf{Inclusion Criteria}} \\
\midrule
I1 & The participant uses AI-generated code in their software projects. \\
I2 & The participant has shared meaningful and consistent responses. \\
\midrule
\multicolumn{2}{l}{\textbf{Exclusion Criteria}} \\
\midrule
E1 & The participant does not use AI-generated code in their software projects. \\
E2 & The participant has shared meaningless and inconsistent responses. \\
\bottomrule
\end{tabular}
\caption{Inclusion and exclusion criteria for collecting survey responses}
\label{tab:surveycriteria}
\vspace{-2.8em}
\end{table}

We included a prerequisite question about whether participants have used \aigc{} in practice. We received 12 responses from practitioners who declared that they do not use any \aigc{} in their projects, and thus, we excluded those responses from our analysis. We also excluded 15 other responses that, upon analysis, were found to be unclear or contained irrelevant answers to the survey questions. %For instance, we excluded a response based on an unclear answer to SQ12, ``\textit{Because it is generated by AI and I cannot declare it as self generated}''. This participant misinterpreted ``self-declared \aigc{}'' as ``self-generated code'', thus providing an unclear response.

%In total, we excluded 15 unclear responses based on this criterion. 

\textit{Analyzing Responses.} %Details of the survey questionnaire are shown in Table \ref{tab:survey}. 
We analyzed all responses to the survey questionnaire in Table \ref{tab:survey} using a combination of quantitative and qualitative analysis methods. For the first nine closed-ended questions, we employed descriptive statistics to ascertain the frequency of various responses, which provides an overview of the participants' demographics and their usage of \aigc{}. For the remaining three open-ended questions (SQ10, SQ11, SQ12), we employed thematic analysis and utilized the inductive approach for analysis~\cite{braun2006using}. Specifically, we first performed initial coding and then constantly compared the codes to develop high-level themes. The first author performed the initial coding by examining each valid response to extract and summarize the core ideas articulated in participants' responses regarding their reasons to self-declare \aigc{}. \textcolor{black}{The initial coding results were shared with the co-authors via a collaborative document. The refinement process took place in two phases. In the first phase, the co-authors reviewed the document and shared their written comments. Afterward, any disagreements were addressed during live meetings.} \textcolor{black}{This approach aligns with established qualitative research practices, where a single primary coder conducts the initial coding while co-authors engage in iterative review and discussion to enhance the reliability of the coding process \cite{McDonald2019reliability}.}

\textcolor{black}{For instance, in the answer to SQ11 about reasons for self-declaring \aigc{},
% commented one example to shorten it, as two examples are enough.
% a participant expressed a positive view of self-declaration, stating in $R_{62}$, ``\textit{It fosters trust with clients and team members by being upfront about the use of AI in the development process}''. This response was coded as \textit{``fostering trust through transparency''}.
a participant emphasized the importance of tracking \aigc{}, stating in $R_{69}$, ``\textit{Placing comments at the exact location of AI-generated code allows for pinpointed identification of machine-generated contributions, which is essential for precise tracking and review}''. This was coded as \textit{``tracking \aigc{} for review''}. In contrast, some participants provided reasons for not self-declaring the use of AI in development in their answers to SQ12. A participant mentioned in $R_{46}$, ``\textit{I'll check the code over myself, the AI is just there to speed things up}'', which we coded as \textit{``do not rely on AI tools but use them just for assistance''}.} 

\textcolor{black}{In this analysis, we performed iterative and constant comparisons to relate and contrast the codes identified from participants' responses. The first author compared the codes derived from specific summarized ideas to identify patterns and similar semantic meanings. 
% For example, responses that reflected trust, transparency, and accountability were grouped into the \textit{transparency, accountability, and ethical considerations} theme. 
Responses that emphasized the need for self-declaring \aigc{} to track and monitor it for future code reviews were categorized into the \textit{Tracking and Monitoring Code for Later Reviews and Debugging} theme. Responses where participants mentioned custom review and modification of \aigc{} as their reasons for not self-declaring it were grouped under the \textit{Customization and Modification} theme. All authors were involved in the validation of the developed codes. In cases of disagreement, all authors engaged in discussions to reconcile differences and reach a consensus on the coding results.} 

\subsection{Mining Study Overview}\label{subsec:miningdemographics}
We collected 613 \aigc{} snippets from 586 GitHub repositories. We gathered information about the collected code snippets to investigate the nature of the repositories to which they belong, including the programming languages and application domains, and to explore the AI code generation tools used. 

\textit{Code generation tools.} Our study focused on four AI code generation tools: \textit{Copilot}, \textit{ChatGPT}, \textit{CodeWhisperer}, and \textit{Code Llama} (see details of the selection process in Section \ref{subsec:Miningstudydesign}). We found that the highest number of \aigc{} snippets are generated by ChatGPT (336, 56.0\%), followed by Copilot (257, 43.0\%). The remaining five snippets (<1\%) are generated by CodeWhisperer. We did not find any code snippets generated by Code Llama, as declared by the developers themselves.

\begin{figure}[h]
    \centering
    \includegraphics[width=0.8\linewidth]{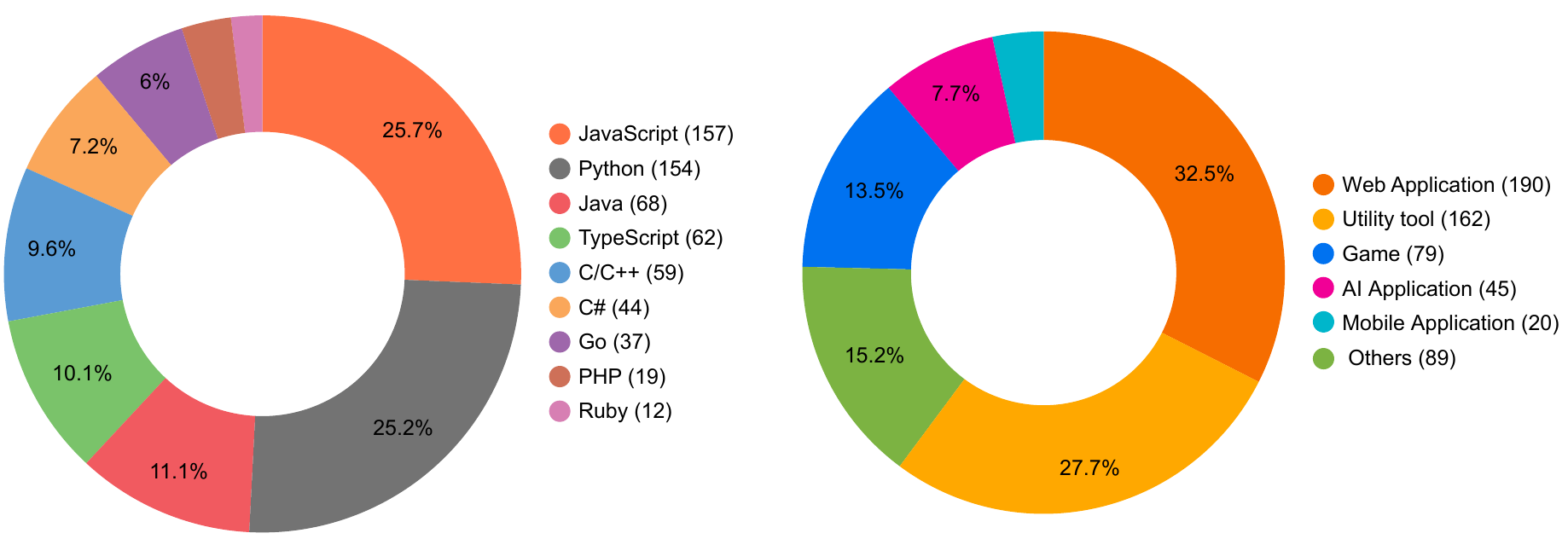}
    \caption{\textcolor{black}{Distribution of collected code snippets across programming languages and application domains}}
    \label{fig:domains}
    \vspace{-1em}
\end{figure} 

\textit{Programming languages.} A summary of the collected code snippets concerning the programming languages used is shown in Figure~\ref{fig:domains}.
We collected self-declared \aigc{} snippets from the 10 most popular programming languages on GitHub. Most of the collected code snippets were written in JavaScript (157, 25.7\%), followed by Python (154, 25.2\%), Java (68, 11.1\%), TypeScript (62, 10.1\%, ), C/C++ (59, 9.6\%), C\# (44, 7.2\%), and Go (37, 6.0\%). %The lowest numbers of code snippets were found in PHP (19, 3.0\%) and Ruby (12, 1.9\%). 

\textit{Application domains.}
We found self-declared \aigc{} in projects belonging to various application domains, with \textit{Web applications} being the most common domain with 191 (32.5\%) repositories, followed by \textit{Utility tools} with 162 (27.7\%) repositories and \textit{Game applications} with 79 (13.5\%) repositories. 89 repositories were not categorized because the project descriptions were either not provided or unclear. Figure \ref{fig:domains} shows a summary of the distribution of self-declared \aigc{} snippets across application domains. 

\begin{figure}[h]
    \centering
    \includegraphics[width=1 \linewidth]{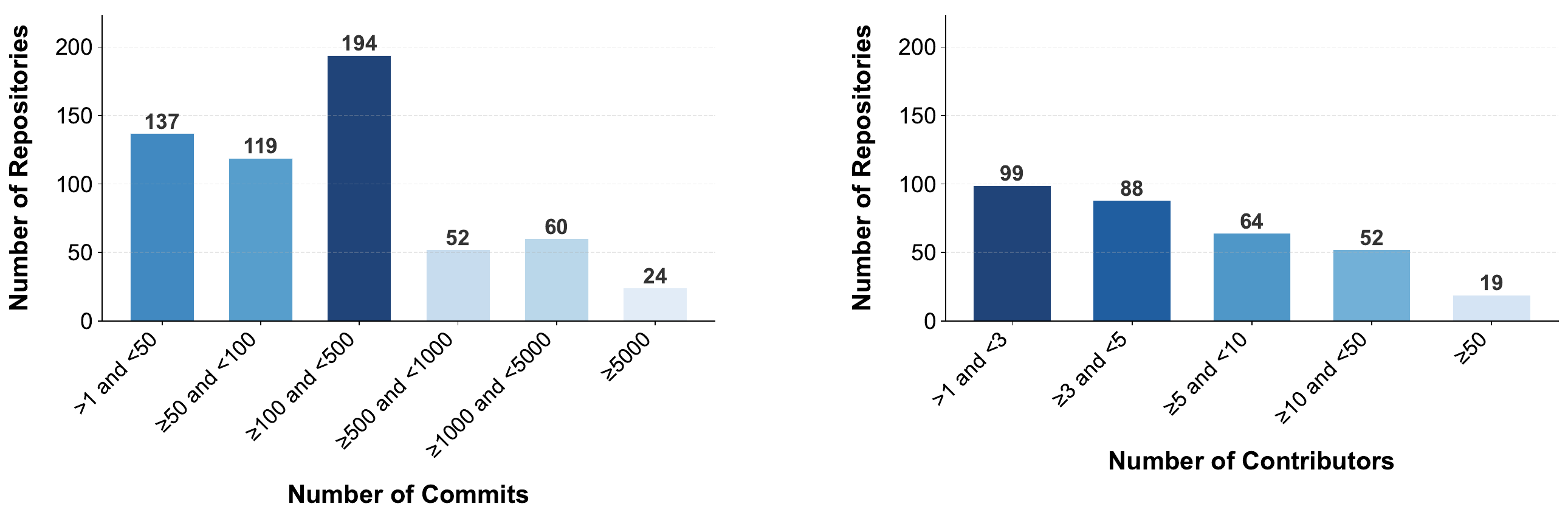}
    \caption{\textcolor{black}{Distribution of repositories based on number of commits and multiple contributors}}
    \label{fig:commits}
    \vspace{-1em}
\end{figure} 

\textit{Commit history.}
Figure \ref{fig:commits} shows the commit history of the repositories that contain self-declared \aigc{}. We categorized repositories into six commit count ranges: from  1 -- 50 to  $\geq$ 5000. In total, we collected data from 586 repositories.  194 (33.1\%) repositories are in the 100 - 500 commits range, indicating reasonable activity in these projects. Moreover, 137 repositories (23.4\%) contain commit counts ranging from 1 to 50, showing certain inactivity in these projects. The projects containing more than 500 commits account for 136 (23.2\%) repositories. 

\textit{Repository contributors.}
Of the 586 repositories, we were able to identify the contributors' information in only 322 of them. The absence of contributor details in the remaining repositories can be attributed to privacy settings (e.g., profiles are not made public) or repositories authored solely by one contributor. As shown in Figure \ref{fig:commits}, these repositories are maintained mostly by small teams. Specifically, 99 of those repositories have between 2 and 3 contributors, while 88 repositories have between 3 and 5 contributors, indicating that small teams maintain 58.0\% of the repositories. Additionally, 64 repositories feature 5 to 10 contributors. Moreover, there are 52 repositories (35.6\%) with 10 to 50 contributors. A relatively small proportion of repositories (19, 11.0\%) have more than 50 contributors.

% \begin{figure}[h]
%     \centering
%     \includegraphics[width=0.8\linewidth]{Figures/contributors.pdf}
%     \caption{Distribution of repositories with multiple contributors}
%     \label{fig:contributors}
%     \vspace{-1em}
% \end{figure} 

\begin{tcolorbox}[
    colback=lightgray!20, 
    colframe=darkgray,   
    boxrule=0.5mm,        
    arc=2mm,              
    title= Summary of Mining Study Overview,      
    fonttitle=\bfseries   
]
The collected code snippets are primarily written using ChatGPT and GitHub Copilot. Almost half of these code snippets were written in JavaScript and Python, and \textit{Web application} was the dominant application domain. The repositories that contain the collected code snippets in our dataset are maintained mainly by small teams (less than five contributors, 58.0\%).
\end{tcolorbox}

\subsection{Survey Demographics}  \label{subsec:surveydemographics} 
\begin{figure}[h]
    \centering
    \includegraphics[width=1\linewidth]{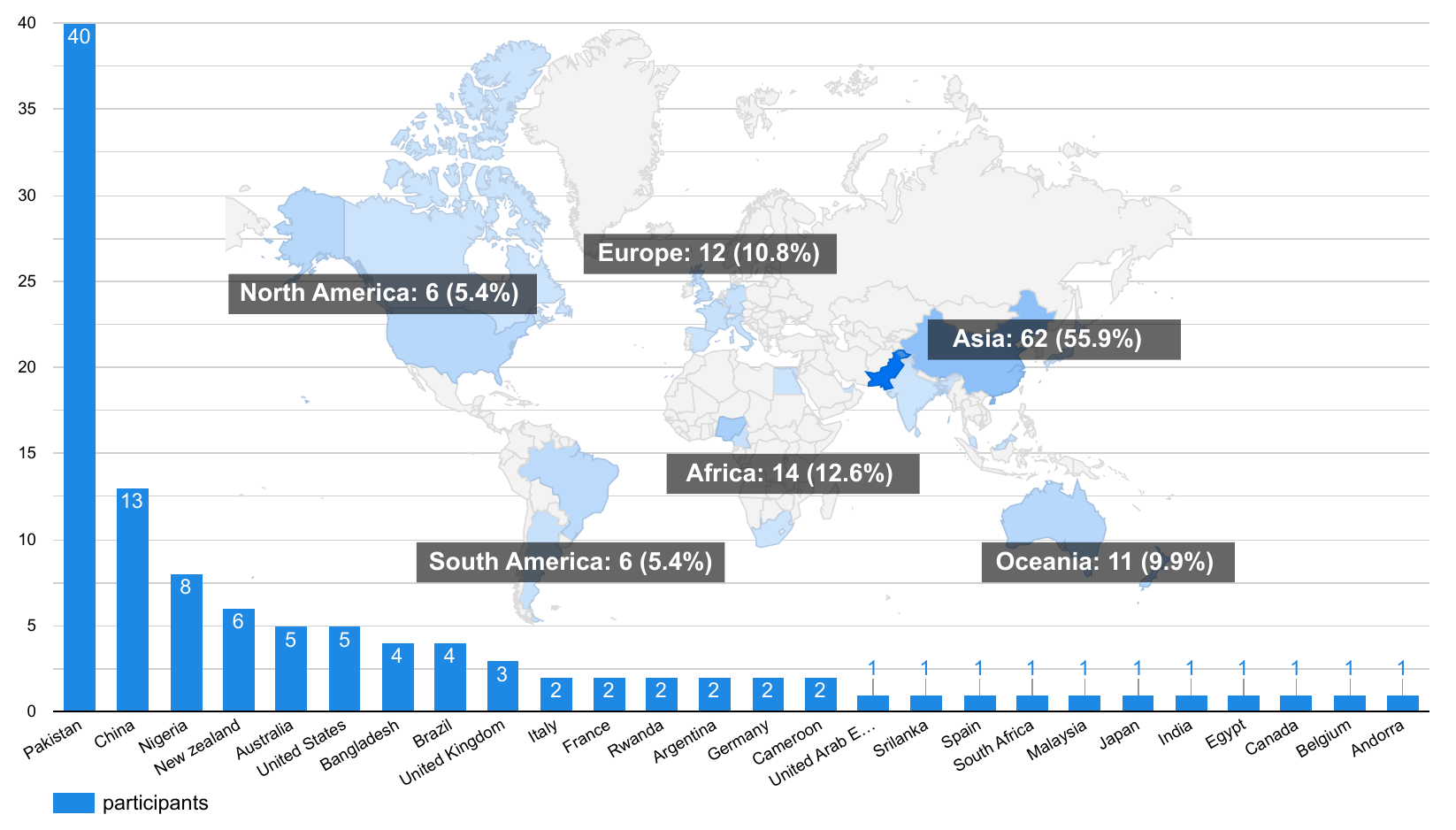}
    \caption{Overview of countries and continents of survey participants}
    \label{fig:countries}
    \vspace{-1em}
\end{figure}

Our survey covers a diverse sample of participants from industry, with 111 valid responses collected from 26 countries, including participants from Pakistan (40, 36.0\%), China (13, 11.7\%), Nigeria (8, 7.2\%) and New Zealand (6, 5.4\%). We received 44 responses from other countries, including Australia, the United States, Bangladesh, Brazil, and the United Kingdom, among others. A summary of participants' location is shown in Figure~\ref{fig:countries}.

\begin{figure}[h]
    \centering
    \includegraphics[width=0.8\linewidth]{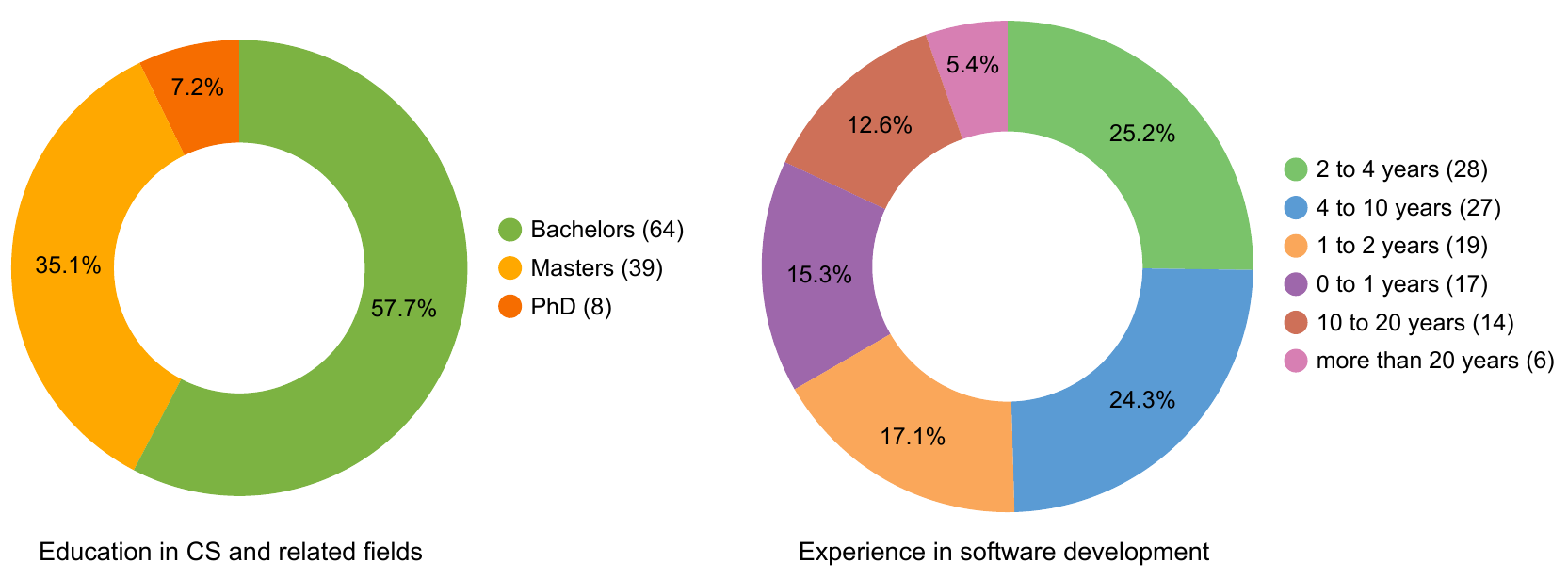}
    \caption{Overview of education and experience of survey participants}
    \label{fig:educationexperience}
    \vspace{-1em}
\end{figure}

\textit{Educational background.} We also analyzed the educational background of all survey participants. All participants hold at least a Bachelor's degree in computer science or related fields. %, which ensures they possess the educational qualifications necessary for our practitioners’ survey. 
Among the participants, %64 hold at leasBachelor's degree, 
39 hold a Master's, while 8 hold doctoral degrees. Figure \ref{fig:educationexperience} shows a summary of the participants' educational backgrounds.

\textit{Experience.} Participants consist of developers with all levels of experience. Nineteen participants have 1 to 2 years of experience, 28 participants have between 2 and 4 years of experience, 27 participants have 4 to 10 years of experience, and 6 participants have over 20 years of experience. A summary of the level of experience is shown in Figure \ref{fig:educationexperience}. 

\begin{figure}[h]
    \centering
    \includegraphics[width=1\linewidth]{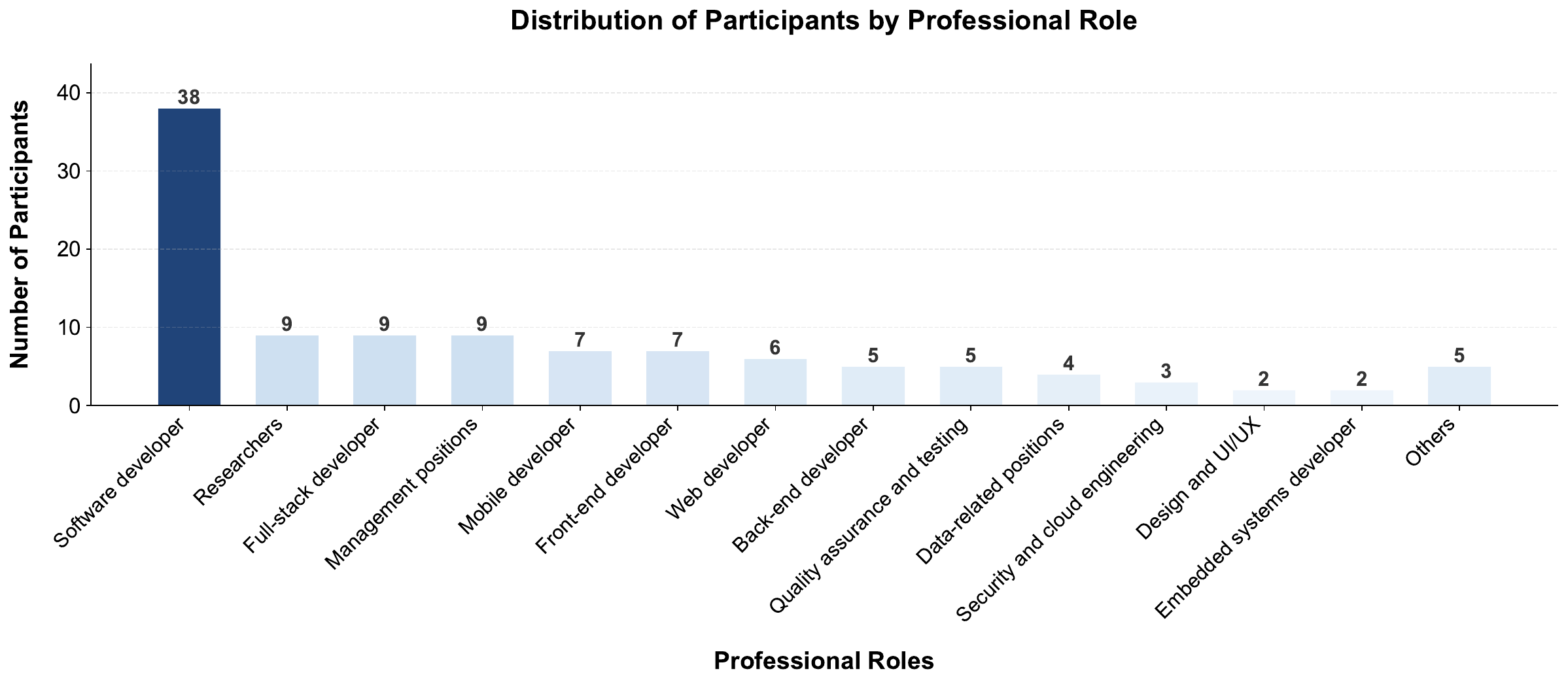}
    \caption{\textcolor{black}{Overview of 111 total survey participants' roles in their organizations}}
    \label{fig:roles}
    \vspace{-1em}
\end{figure}

\textit{Professional roles.} Most participants (74, 67.2\%) reported that they work in software development roles in their organizations. \textcolor{black}{38 participants provided a broad development role ``\textit{software developer}'' while 36 participants mentioned specialized roles, including full-stack developer, backend developer, and mobile developer in their responses}. Participants also work in management positions, including team leads, engineering managers, and CTOs (9 participants), as well as researchers (9 participants). Other roles represented in the survey include quality assurance and testing (5 participants), data-related positions (4 participants, e.g., data analysts and database administrators), security and cloud engineering (3 participants), and design and UI/UX (2 participants). An overview of the participants' roles in their organizations is shown in Figure~\ref{fig:roles}.

\textit{Application domains.} The participants work in various application domains. The largest group, comprising 82 participants) works mainly in Web applications, followed by 51 participants who work in Mobile applications, 40 in AI applications, 33 in utility tools, and 12 in games.

\textit{AI tools used.} The most commonly used AI code generation tool noted by participants was ChatGPT (102), followed by Copilot (52), Claude (28), CodeWhisperer (12), and Gemini (3).

\begin{tcolorbox}[
    colback=lightgray!20, 
    colframe=darkgray,   
    boxrule=0.5mm,        
    arc=2mm,              
    title= Summary of Survey Demographics,      
    fonttitle=\bfseries   
]
Our survey received 111 valid responses from 26 countries. The survey participants mostly work in the industry in software development-related roles. The majority of survey participants work on web applications and prefer \textit{ChatGPT} and \textit{GitHub Copilot} for generating code.  
\end{tcolorbox}

\section{Results}\label{sec:results}

\subsection{RQ1: How do developers self-declare \aigc{} in their projects?} \label{subsec:resultRQ1}

\subsubsection{RQ1.1: What information do developers share in their \sdc{}?}
\begin{figure}[h]
    \centering
    \includegraphics[width=1\linewidth]{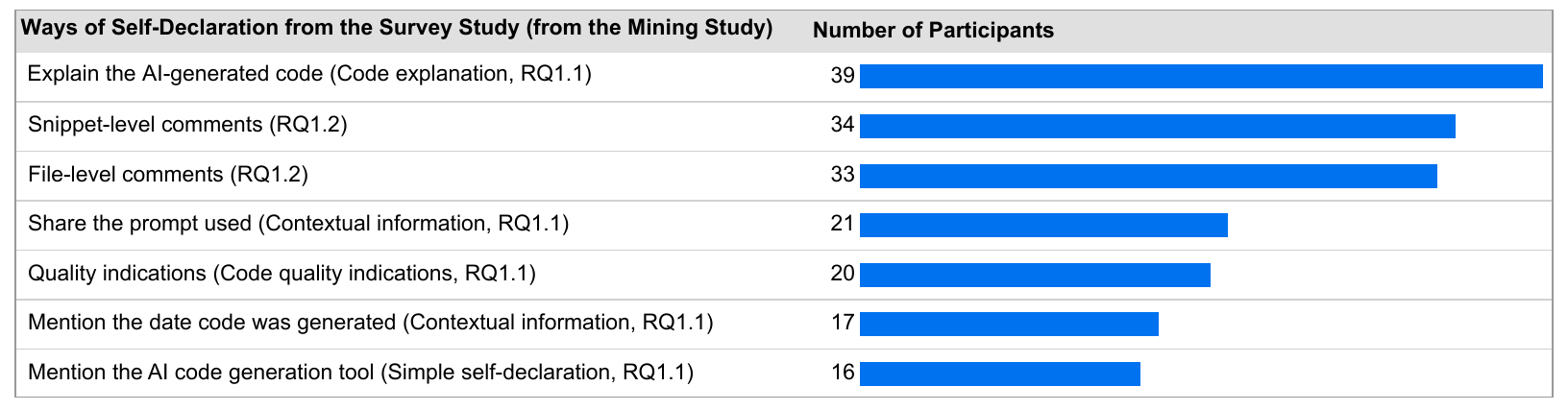}
    \caption{Ways of self-declaration comments from the survey study and their mappings to the ways from the mining study}
    \label{fig:practices}
    \vspace{-1em}
\end{figure}

We extracted \sdc{} (i.e., comments associated with the code that helped us identify if the code was AI-generated) from the collected \aigc{} snippets. We present the categories of the \sdc{} and related examples of such comments in Table \ref{tbl:Comments}. These categories show that the common ways developers write \sdc{} are: 1) \textit{simple self-declaration} (simple comment mentioning that this code is generated by AI), 2) \textit{code explanation} (a brief description of \aigc{}), 3) \textit{contextual information} (context of code generation such as prompts given to the AI code generation tool), and 4) \textit{code quality indications} (include observations about the quality of the \aigc{}). We discuss each category below. The categories identified in the mining study and the survey study, and their mapping relationships, are presented in Figure \ref{fig:practices}.  

1) \textit{Simple self-declaration.} We identified 214  (34.9\%) \sdc{} as \textit{simple \sdc{}}. These comments self-declare \aigc{} in a few words, simply acknowledging the code is AI-generated. In most cases (over 96\%), the comment will also include the name of the code generation tool used. The example (Comment 1) in Table \ref{tbl:Comments} shows a developer mentioning that Copilot was used to generate the code. Only 3.3\% of these comments just self-declared the \aigc{} without mentioning the tool name. For example, we can see in Comment 3 in Table \ref{tbl:Comments} that a developer simply self-declared a function as AI-generated without mentioning any further details. Some \sdc{} (4\%) also acknowledged the simultaneous use of multiple AI code generation tools. In Comment 2 in Table \ref{tbl:Comments}, a developer mentioned GitHub Copilot and ChatGPT simultaneously for code generation.  

The survey participants also indicated that they used simple self-declaration for \aigc{}. Sixteen participants (14.4\%) reported that they simply provided the name of the AI code generation tool used for code generation in the self-declaration comment. %(see Figure \ref{fig:practices}). 
Participants mostly shared reasons for using simple self-declaration comments to track and distinguish AI-generated code from human-written code. In $R_{28}$, a participant mentioned that ``\textit{I only need to distinguish `my code' from AI-generated code}''. 

\begin{table}[]
\footnotesize
\setlength{\tabcolsep}{5pt}
\renewcommand{\arraystretch}{1.3}
\begin{tabular}{>{\raggedright\arraybackslash}p{0.15\linewidth}>{\raggedright\arraybackslash}p{0.62\linewidth}>{\raggedright\arraybackslash}p{0.13\linewidth}}
\toprule
\textbf{Categories}                & \textbf{Self-Declaration Comment}           & \textbf{Percentage (\%)} \\
\midrule

\multirow{3}{=}{Simple Self-declaration} 
& 1. This function was mostly written by Copilot (\href{https://github.com/bmschmidt/wax-kit/blob/906468ad6cc99e65a0cc1f4aa03b566cd37eb651/src/lib/image_management.js#L24}{comment})  & \multirow{3}{*}{(214) 34.9\%} \\
& 2. Coded by [developer name] and GitHub Copilot (and ChatGPT) (\href{https://github.com/MirkoTh/exploration-psychometrics/blob/c155168570f352129e6703880845c41e295a1f19/task/task.js#L3}{comment})  & \\
& 3. Note: AI-generated function (\href{https://github.com/rviscomi/capo.js/blob/84691c922b1a23991b89c937c80ca32acab87fbc/src/lib/io.js#L94}{comment})  & \\
\hline

\multirow{2}{=}{Code Explanation} 
& 4. Function that creates lighter shades of a color, written by Copilot (\href{https://github.com/khaihernlow/santa-tracker-frontend/blob/4d411b5c1cb1c362830b488699ecfb15da68cb23/js/santaController.js#L364}{comment}) & \multirow{2}{*}{(185) 30.0\%} \\
& 5. Grayscale the entire page except the specified element solution by ChatGPT (\href{https://github.com/ThibaudLopez/chrome_resizer/blob/761e4e79daf9dd067f04deae3953551119a49e87/content.js#L46}{comment}) & \\
\hline

\multirow{2}{=}{Contextual Information} 
& 6. Function written thanks to ChatGPT (\href{https://chat.openai.com/share/00e006b8-71f3-43d0-80dc-774c4551f6e3}{ChatGPT conversation link}) (\href{https://github.com/diamondburned/bulbremote/blob/5ce7fdcc6ebb1abb790afdcf239760994cf9939e/src/lib/tasmota.ts#L62}{comment}) & \multirow{2}{*}{(209) 34.0\%} \\
& 7. Written by [developer name], 4/9/2024 with the assistance of GitHub Copilot (\href{https://github.com/LCC-CIT/CS233JS-BuncoGame/blob/e0bf7624be299ed1bca3e4e23006b6c1d97c7d53/scripts/die.js#L1}{comment})  & \\
\hline

\multirow{2}{=}{Code Quality Indications} 
& 8. Player Class Entirely written by humans with assistance from Copilot (75\% human, 25\% AI) (\href{https://github.com/Alexandre2006/PlatformGame/blob/f17b38a107087f8a90f48192c2362624bf3123e0/utils/player.py#L2}{comment}) & \multirow{2}{*}{(110) 18.0\%} \\
& 9. \# NOTE: (11-24-2023) most of this code was generated by ChatGPT (GPT-4). It is a quick solution and needs refinement (\href{https://github.com/anarchy-ai/llm-speed-benchmark/blob/2b2917c390074fba879bf5090139960d51999561/tools/graph.py#L1}{comment}) & \\
\bottomrule
\end{tabular}
\caption{\textcolor{black}{Ways and their examples of self-declaration comments from the mining study}}
\label{tbl:Comments}
\vspace{-3.4em}
\end{table}

2) \textit{Code explanation.} We also found comments (185, 30.0\%) where developers shared the explanation of \aigc{}. These comments explain how the code is generated, what it is about, and how it works. In most cases, these code explanations are brief (83.0\% of the 185 comments), typically one sentence long; however, some comments contain more significant details (17.0\%). In the example shown in Table \ref{tbl:Comments} (Comment 4), a developer self-declared the function \texttt{shadeColor(color, percent)} as \aigc{} and also explained that its functionality is to generate a lighter shade of a color. Similarly, in Comment 5, a developer self-declared and explained the function \texttt{grayscale\_all\_but(element)} that is used to grayscale the entire page, except for the element passed as a parameter. 

The survey results further confirm these results. We found that 39 participants (35.2\%) cited that they provide code explanations when they self-declare their \aigc{}. Participants reported that they typically include text that helps other developers understand the added code. For example, one of the participants explained their reason for writing code explanations in their \sdc{} in $R_{19}$, ``\textit{I like to explain the \aigc{} because it helps others understand what it does and why it’s there}''. 

3) \textit{Contextual information.} We found 209 (34.0\%) \sdc{} where the details provided additional contextual information about the added code, which includes information such as the author's name, the date the code was written, the version of the AI tool, and the prompts used to generate the code. For example, Comment 6 shown in Table \ref{tbl:Comments} shows a developer who shared the link to the ChatGPT conversation on generating the function \texttt{cct2chs(w: number, s: number, v: number)}. Similarly, we can see that a developer added their username and the date the code was generated in another example (Comment 7 in Table \ref{tbl:Comments}).

Several survey participants also expressed a similar view as they also include contextual information about the code when they self-declare \aigc{}. Twenty-one participants (18.9\%) mentioned that they provided the prompt to generate the code in the self-declaration comment, with a further 17 participants (15.3\%) also reporting that they also provided the date the code was generated.

4) \textit{Code quality indications.} Several \sdc{} include a comment about the quality of the generated code, in both positive and negative tones. 48 (8.0\%) of comments show that the \aigc{} was modified by a human to transform it for specific requirements. This shows the developer's contribution to the code, a positive sign for code quality. Some of these comments also mentioned that the generated code was tested and worked per their use case. For instance, in Comment 8 in Table \ref{tbl:Comments}, a developer self-declared ``\textit{player}'' class and specified that only 25.0\% of the code is generated using AI and 75.0\% is the human contribution, which is a positive indication. On the contrary, 62 (10.0\%) comments highlighted the possibility of errors in the \aigc{} and asked for additional testing and improvements. This type of self-declaration comment shares negative indications about the quality of \aigc{}. Comment 9 in Table \ref{tbl:Comments} shows an example of such comments. In this example, a developer admitted that they provided a quick solution that requires refinements in the self-declaration comment, which is a negative sign for code quality.  

Similarly, in our survey, 20 participants (18.0\%) reported including a note about the code quality when self-declaring \aigc{}. A participant explained that the reason for including code quality indications in their \sdc{} is ``\textit{to keep track of whether the code is modified by a human person and tested according to the requirements of project}'' ($R_{10}$). 

\begin{tcolorbox}[
    colback=lightgray!20, 
    colframe=darkgray,   
    boxrule=0.5mm,        
    arc=2mm,              
    title=Finding 1,      
    fonttitle=\bfseries   
]
Developers employ various methods to self-declare AI-generated code in their projects, including simply stating the use of AI in a single sentence, providing a brief explanation of the code, sharing contextual information (such as the prompts used to generate the code), or reporting the code quality either positively or negatively.
\end{tcolorbox}

\subsubsection{RQ1.2: What is the scope of the \sdc{}?} \label{subsubsec:resultRQ1.2}
We identified two main \sdgc{} scopes: \textit{\snl{}} and \textit{\fl{}}. 
% Note that not all participants selected the scope in the SQ9. %The answers to this RQ provide the scope of the \sdc{}, i.e., self-declaration of a code snippet or a whole file, when they appear in the source code files. 

\begin{figure} [h]
    \centering   \includegraphics[width=1 \linewidth]{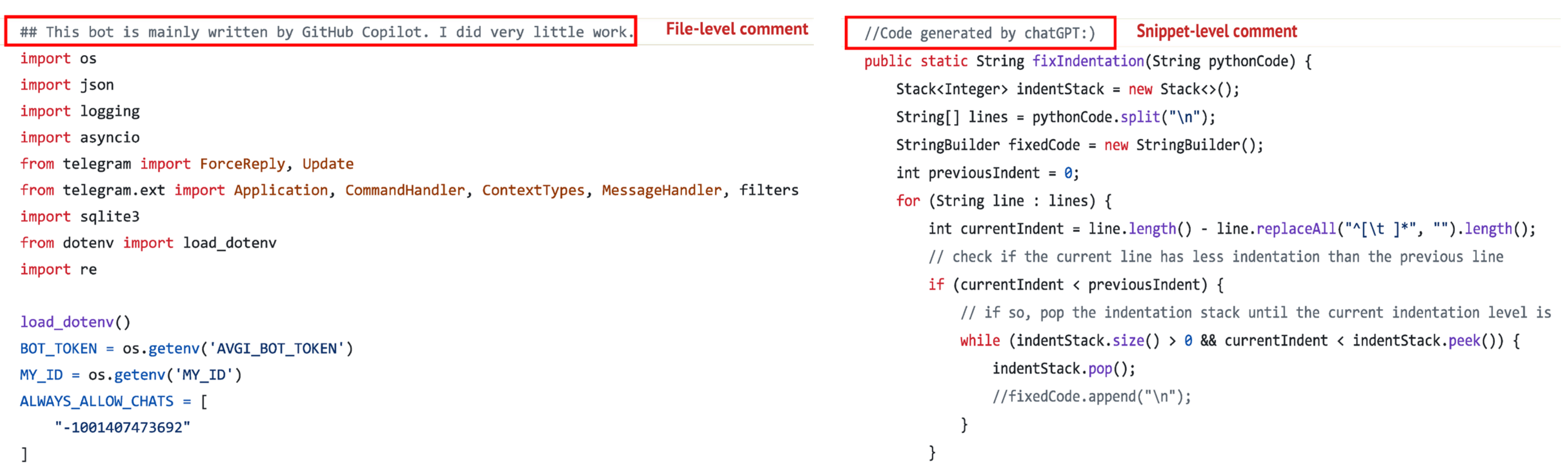}
    \caption{\textcolor{black}{Examples of \fl{} and \snl{} self-declaration comments}}
    \label{fig:filesnippet}
    \vspace{-1em}
\end{figure}

% \begin{figure} [h]
%     \centering   \includegraphics[width=0.8\linewidth]{Figures/file-level-1.pdf}
%     \caption{An example of \fl{} self-declaration comment declaring the entire file as mostly AI-generated}
%     \label{fig:file1}
%     \vspace{-1em}
% \end{figure}

\textit{File-level comments.} In the \fl{} \sdc{}, a developer self-declares the entire file as AI-generated without specifying the exact location within the file. These comments have two possible cases: 1) the entire file is \textit{mostly} AI-generated with little or no human contribution, or 2) the specific code file contains \textit{some} \aigc{}, not mentioning the exact location of the code snippet. These comments are typically added at the \textit{start} or \textit{end} of the source code file. An example of a \fl{} \href{https://github.com/ksqsf/toys/blob/cd2fe7ef30d29e88c02b6d83cf43f3a4a182c441/telegram/logger.py#L1}{comment} of the first case is shown in Figure \ref{fig:filesnippet}. The developer mentioned at the beginning of the file that the added code is mostly AI-generated. %\textit{``This bot is mainly written by GitHub Copilot. I did very little work''}. 
Therefore, the scope of this self-declaration comment is the entire source code file. In total, we found 187 (30.5\%) \fl{} \sdc{}. 

The responses to the practitioners’ survey show similar results, with 33 participants (29.7\%) indicating that they use \fl{} comments to self-declare their \aigc{}. A participant explained when they employ \fl{} comments: ``\textit{I use \fl{} comments when I use intensively integrated \aigc{}, and I use snippet-level comments when integrate less \aigc{}}'' ($R_{10}$).

\textit{Snippet-level comments.} Developers include these comments at specific locations of the source file (such as before a method/function or a class) to indicate that this particular portion of the code is AI-generated. For example, if a developer adds a comment above a function, it suggests that the comment pertains to that specific function.
Figure \ref{fig:filesnippet} shows an example of \snl{} \href{https://github.com/albilu/netbeansPython/blob/764d405388b1921f6248a9782cdcae78c1b06f17/src/main/java/org/netbeans/modules/python/actions/PythonFixIndentAction.java#L49}{comment}. In this example, the developer self-declared the \aigc{} \textit{before} the \texttt{fixIndentation()} function. This suggests that this self-declaration comment applies \textit{only} to the function. %We classified such cases as \snl{} comments, as their scope is limited to a particular code snippet that follows. 
Developers self-declare various code sections in \snl{} comments, including methods/functions, classes, conditions, loops, or statements. Our findings indicate that developers primarily self-declare \aigc{} at the class and function level. We found 426 instances of \snl{}{} self-declaration comments, accounting for 69.4\% of all collected comments.

\begin{figure} [h]
    \centering   \includegraphics[width=0.9 \linewidth]{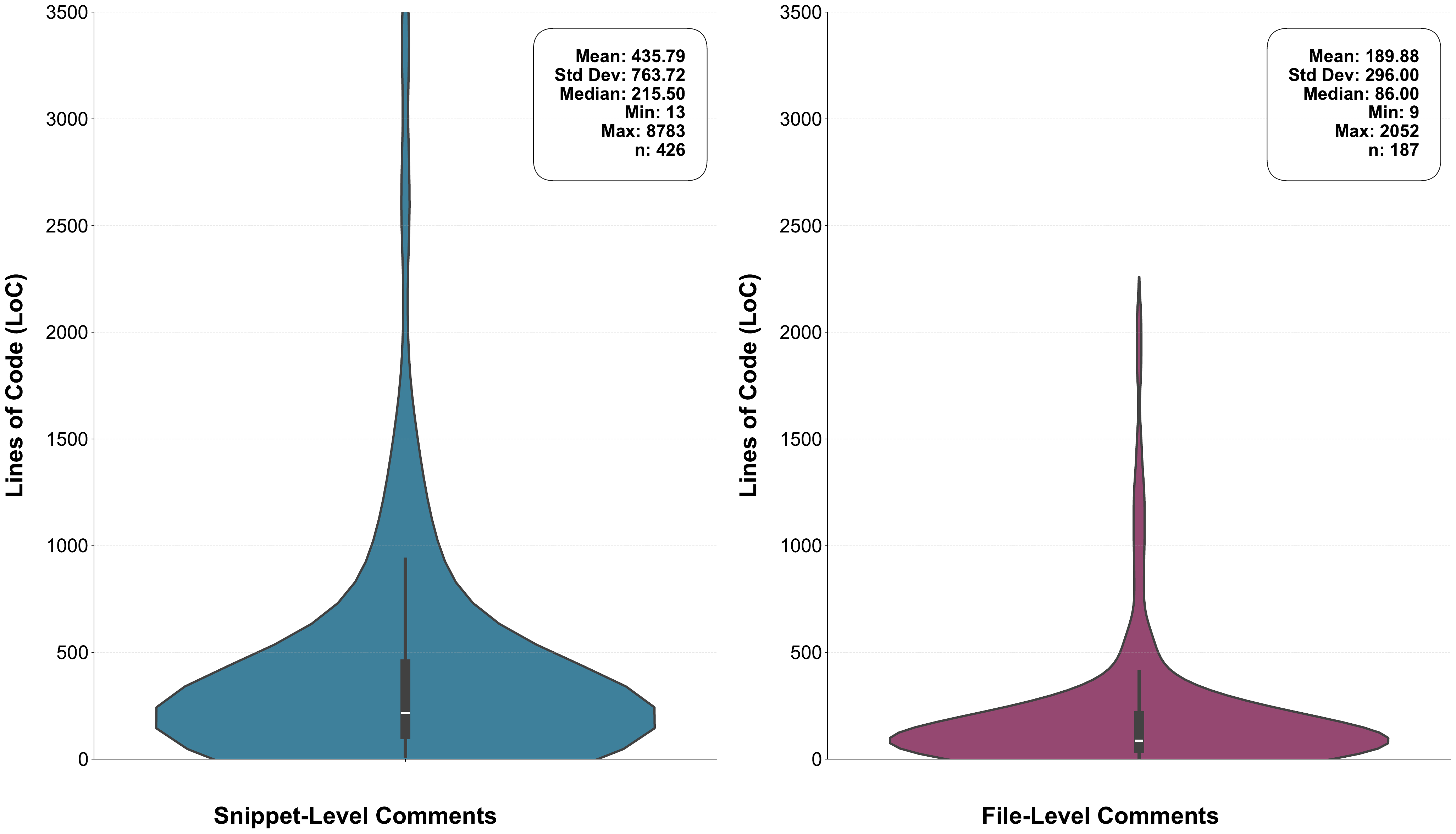}
    \caption{\textcolor{black}{LoC of files containing \fl{} and \snl{} self-declaration comments}}
    \label{fig:filesnippetloc}
    \vspace{-1em}
\end{figure}

In the practitioners’ survey, 34 participants (30.6\%) mentioned that they utilize \snl{} \sdc{} to self-declare their \aigc{}, and reported the reasons such as showing exactly which part of the code is AI-generated as shared by $R_{19}$, ``\textit{I use \snl{} comments to point out the AI-generated parts because it gives clear context right where it’s needed. This way, anyone looking at the code can easily see what was generated by AI}''. 

\textcolor{black}{We also calculated the lines of code (LoC) of all 613 code files in our dataset. Then, we compared the LoC of files containing snippet-level comments with files containing file-level comments to understand their relationship with file size. Figure \ref{fig:filesnippetloc} demonstrates that snippet-level comments are found in larger code files compared to file-level comments found in shorter code files. The statistical analysis of LoC data reveals a substantial difference in file sizes: files with snippet-level comments have a mean of 436 LoC (median 215), while files with file-level comments average 190 LoC (median 86).  Although there is high variability in file sizes for snippet-level comments, 54\% of the files are over 200 LoC. On the contrary, 75\% of the files containing file-level comments are below 200 lines of code. This significant difference in the size of files with snippet-level and file-level comments suggests that developers tend to use snippet-level comments for larger, more complex files and file-level comments for shorter code files.}

% Many comments also included \aigc at the class level, but conditions and loops were rarely self-declared in the \snl{}{} comments. 

% \begin{figure}
%     \centering
%     \includegraphics[width=0.8\linewidth]{Figures/snippet-level-2.pdf}
%     \caption{An example of \snl{} self-declaration comment declaring a function as AI-generated}
%     \label{fig:snippet}
%     \vspace{-1em}
% \end{figure}

\begin{tcolorbox}[
    colback=lightgray!20, 
    colframe=darkgray,   
    boxrule=0.5mm,        
    arc=2mm,              
    title=Finding 2,      
    fonttitle=\bfseries   
]
Developers use \textit{snippet-level} comments more often than \textit{file-level} comments to self-declare \aigc{}. When the code file has a few AI-generated code snippets, developers prefer a \textit{snippet-level} comment above each \aigc{} snippet. However, if the code is extensively AI-generated, they tend to write a single \textit{file-level} comment at the beginning or end of the code file.
\end{tcolorbox}

\subsubsection{RQ1.3: What is the distribution of the \aigc{} snippets with \sdc{} across repositories?} 
\begin{figure}[h]
    \centering
    \includegraphics[width=0.9\linewidth]{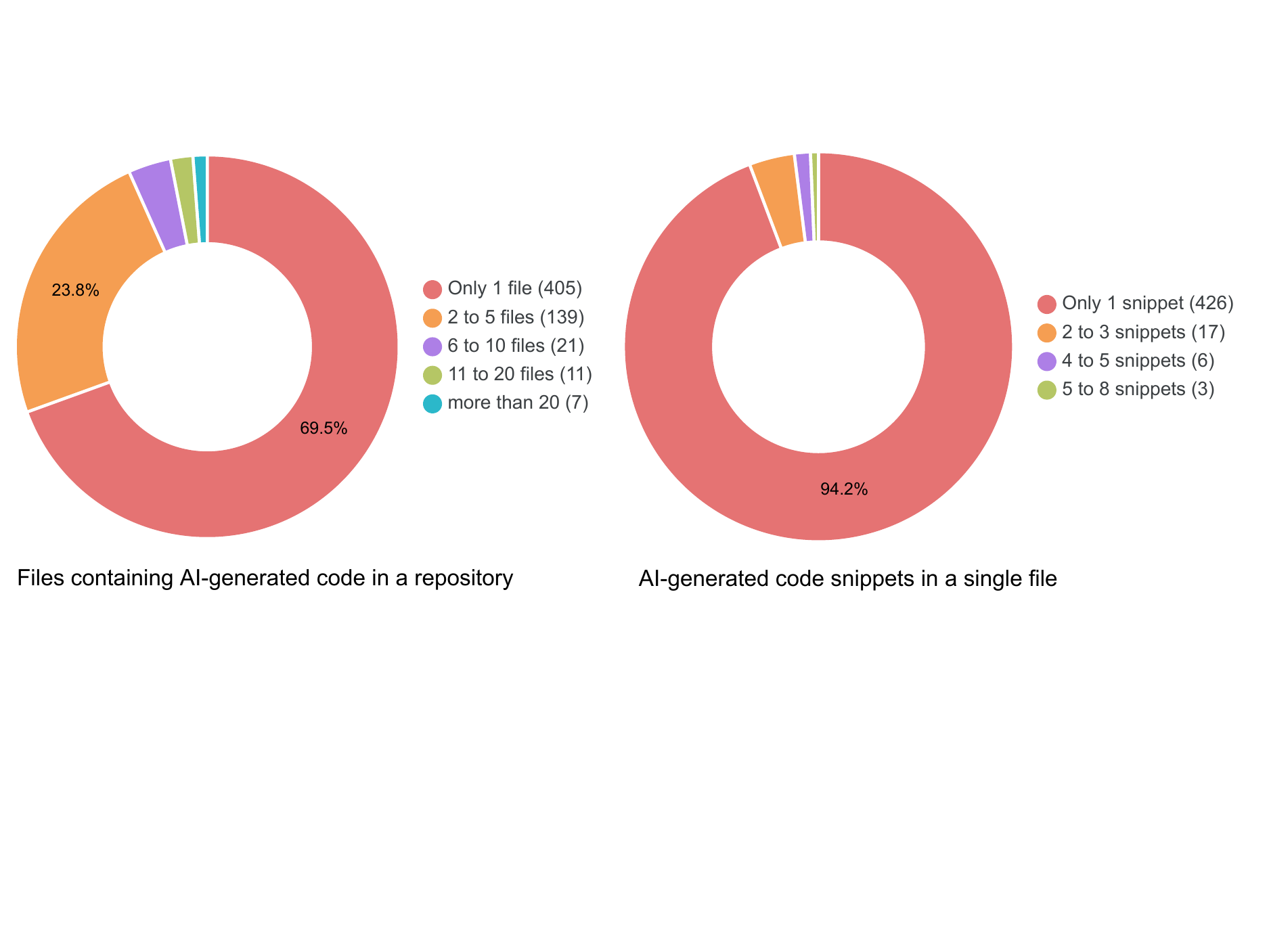}
    \caption{Distribution of \aigc{} in repositories and single files}
    \label{fig:distribution}
    \vspace{-1em}
\end{figure}

We presented the distributions of \aigc{} in repositories and individual source code files in Figure \ref{fig:distribution}. \textcolor{black}{In this section, we refer specifically to self-declared \aigc{}, acknowledging that the actual proportion of AI-generated code in repositories may be higher.}

\textit{Files containing \aigc{} in a repository.} Out of a total of 586 repositories from our dataset, 405 repositories (69.5\%) contain only a single file with \aigc{} snippets. We found 178 repositories (30.5\%) with multiple files containing \aigc{} snippets, with 139 repositories (23.8\%) containing between 2 and 5 files with \aigc{}, while 39 repositories (6.7\% ) with more than five files contained AI-generated code.

\textit{\aigc{} snippets in a single file.} We identified 613 files containing \aigc{} snippets. As presented in Section \ref{subsubsec:resultRQ1.2}, most \sdc{} (426, 69.5\%) are at the \snl{}. 400 out of 426 (93.9\%) of these files have a single \snl{} self-declaration comment. Out of the 426 files containing \snl{} comments, only 26 (6.1\%) have multiple \sdc{} in a single file. The highest number of self-declarations in a single file is 8 \sdc{}. %with an average of one \sdc{} per file.

\begin{figure}[h]
    \centering
    \includegraphics[width=1 \linewidth]{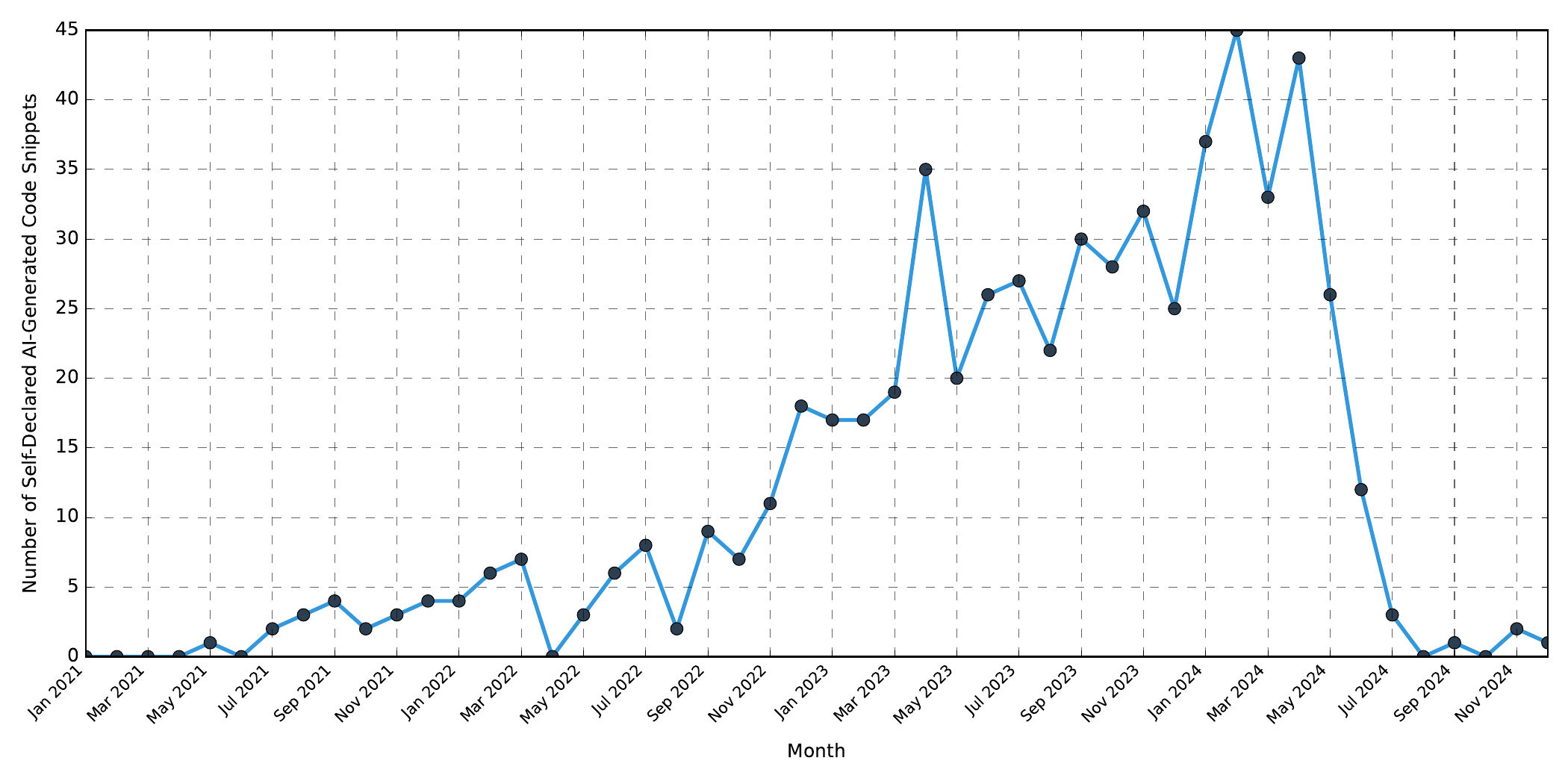}
    \caption{\textcolor{black}{Monthly trend of self-declared AI-generated code spanning Jan 2021 to December 2024}}
    \label{fig:sdc_trend}
    \vspace{-1em}
\end{figure} 

\textit{\textcolor{black}{Trend of self-declaring \aigc{} among developers.}} We visualized the dates when self-declaration comments were added to the files, covering the period from January 2021 to December 2024. We observed an increasing trend in self-declaring AI-generated code among developers, as shown in Figure \ref{fig:sdc_trend}. The trend exhibits continuous growth, punctuated by sudden spikes. A noticeable increase is observed starting in November 2022, coinciding with the introduction of the GPT-3.5 model. Similarly, we observed another significant rise in March 2023, when the GPT-4 model was launched alongside its integration with Copilot X. Additionally, there was a further spike in November 2023, following the release of GPT-4 turbo. Overall, the trend diagram illustrates that the practice of self-declaring AI-generated code is on the rise as more advanced models are introduced. \textcolor{black}{However, we observed a sudden drop in the trend after April 2024. This decline may be due to a few factors: (1) the timing of our data collection from GitHub, as newer self-declared AI-generated code snippets might not have been fully indexed or gained enough visibility to be included in our analysis; and (2) the increasing popularity of other AI code generation tools, including autonomous coding agents, which were not part of our search terms.}

% However, we see a sudden drop in the trend after April 2024, which may be attributed to: (1) the timing of our data collection from GitHub. Newer self-declared AI-generated code snippets may not have been fully indexed on GitHub yet or may not have gained sufficient visibility to be captured in our analysis. (3) The rise in popularity of other AI code generation tools also including autonomous agent tools which are not included in our search terms. }

\begin{tcolorbox}[
    colback=lightgray!20, 
    colframe=darkgray,   
    boxrule=0.5mm,        
    arc=2mm,              
    title=Finding 3,      
    fonttitle=\bfseries   
]
Most of the GitHub repositories in our collected dataset have a few code files with AI-generated code snippets. These code files with \aigc{} snippets typically contain a single self-declaration comment, with only a few code files having multiple \sdc{}. Overall, the occurrences of self-declaration comments in GitHub repositories are low. \textcolor{black}{The trend indicates a steady increase in self-declaring AI-generated code among developers.}
\end{tcolorbox}

\subsubsection{RQ1.4: Are developers satisfied with the \aigc{} according to \sdc{}?}
We analyzed developers' sentiments concerning their added \aigc{}. This analysis provides a general overview of developers' satisfaction with using \aigc{}. Through our manual labeling (see Section \ref{subsec:Miningstudydesign}), we identified 165 (26.9\%) comments that have been labeled as either positive or negative. Overall, we found more positive comments (85.0\%) than negative ones (15.0\%). 

\textit{Developer's satisfaction.} In the positive comments, developers acknowledged the usefulness of the AI code generation tools. We provide two examples of such comments with positive sentiments below. In the first example, a developer self-declared a line of code \texttt{whiteCell = (x // 5) \% 2 != (y // 5) \% 2} as AI-generated and acknowledged the help of the code generation tool: ``\textit{Thanks ChatGPT, I was stuck on this for 2 hours}'' (\href{https://github.com/Kashifraz/SDAGC/blob/main/Mining%20Study/python/PicrossWidget.py#L58}{comment}). In the second example, ChatGPT helped the developer to write a CSS code file, which the developer acknowledged positively:``\textit{github copilot makes it into a real CSS file for me thanks}'' (\href{https://github.com/Kashifraz/SDAGC/blob/main/Mining%20Study/python/main-6.py#L585}{comment}).

\textcolor{black}{We also found reasons for developers' satisfaction that they shared in their positive self-declaration comments. The main reasons we identified are that AI code generation tools save developers time and provide working solutions, help them solve complex logical challenges, assist them in improving the code, and identify critical errors in code. For example, a developer appreciated ChatGPT for helping them generate working code in less time, ``\textit{Btw, everything in class LatencyMonitor (including the class and method docstring comments) was written by ChatGPT. It took about 10-15 minutes of interactive code review between me and ChatGPT to get to this point. Amazing\!!}'' (\href{https://github.com/lil-lab/cb2/blob/72e7ded9ea9908bd0fa259617caaec601fcd637c/src/cb2game/server/util.py\#L194}{comment}). Similarly, a developer acknowledged the help of ChatGPT in identifying the problem in code and improving it, ``\textit{makes sure the user only sees relevant options (only for languages so far), note: ChatGPT helped make this way more efficient - was querying the db too much}'' (\href{https://github.com/jonathanhouge/ReeldIn/blob/4f08c2ab65b7b421d3cc1833405121e854cecdc4/recommendations/helpers.py\#L52}{comment}).}

\textit{Developer's dissatisfaction.} We have also found negative comments where developers provided warnings and disclaimers that the code is generated using AI code generation tools. This can indicate that the generated code may contain bugs or may not meet the quality expectations of developers. Some examples of such comments with negative sentiments are: ``\textit{error handling, really poorly written code, GitHub copilot did it}'' (\href{https://github.com/Kashifraz/SDAGC/blob/main/Mining%20Study/typescript/index-2.ts#L104}{comment}) and ``\textit{disclaimer! This code was written by ChatGPT (with minimal editing) and may not necessarily be the best solution for the problem}'' (\href{https://github.com/Kashifraz/SDAGC/blob/main/Mining%20Study/typescript/removeString.ts#L1}{comment}). Although the developers have integrated these \aigc{} snippets in both examples, they were not quite satisfied with the code quality.
%In the first example, the developer generated \texttt{handleError} function, which is an error handler. Similarly, in the second example, the developer generated \texttt{removeString} function, which removes the first occurrence of string a from string b if a is a substring of b.  These examples highlight the developers' dissatisfaction with the \aigc{}.

\textcolor{black}{We also identified reasons for developers' dissatisfaction shared in their negative self-declaration comments. The main reason for their dissatisfaction is the poor code quality of the AI-generated code. The code quality issues are related to readability and maintainability, code optimization and efficiency, and error handling. For instance, a developer shared that they do not trust the code generated by Copilot because it produced inefficient code, ``\textit{I wrote this with Github Copilot, don't trust it ... Probably uses \*tons\* of API calls. TODO: Can probably be optimized by first getting all the issues/PRs/comments, putting them in some cache, and then iterating over them}'' (\href{https://github.com/ActivityWatch/contributor-stats/blob/95d4eb32b2f43776d1772f411a9af8b2e1abd258/src/contributor_stats/github_stats.py\#L134}{comment}). Similarly, another developer indicated code readability and maintainability issues in the Copilot-generated code, ``\textit{Please excuse the comments and lengthy code, this was written using Copilot and I have not yet cleaned it up}'' (\href{https://github.com/Zay-Codes-Lab/flow-sight/blob/88bb622a2abc8ecf8c8d3dbdcee0fa215d31699b/src/parser/index.js\#L5}{comment}).}

\begin{tcolorbox}[
    colback=lightgray!20, 
    colframe=darkgray,   
    boxrule=0.5mm,        
    arc=2mm,              
    title=Finding 4,      
    fonttitle=\bfseries   
]
Developers' sentiments about the \aigc{} and tools used can be identified in some of the added \sdc{}. The comments are mostly positive, acknowledging the usefulness of the AI code generation tools. However, developers also sometimes express negative sentiments, indicating their dissatisfaction with \aigc{}.
\end{tcolorbox}
 
\subsection{\textcolor{black}{RQ2: What proportion of developers self-declare \aigc{} in practice? Why or why not?}}

\subsubsection{RQ2.1: How common do developers self-declare \aigc{} in their projects?} 
The vast majority of the 111 survey participants (90.2\%) reported actively using AI code generation tools in their development processes. Since our study focuses explicitly on the practices of self-declaring \aigc{}, we excluded responses (12 in total) where developers indicated that they do not actively use AI code generation tools in practice (see Section \ref{subsection:surveydataanalysis}).

\begin{figure}[h]
    \centering
    \includegraphics[width=0.45\linewidth]{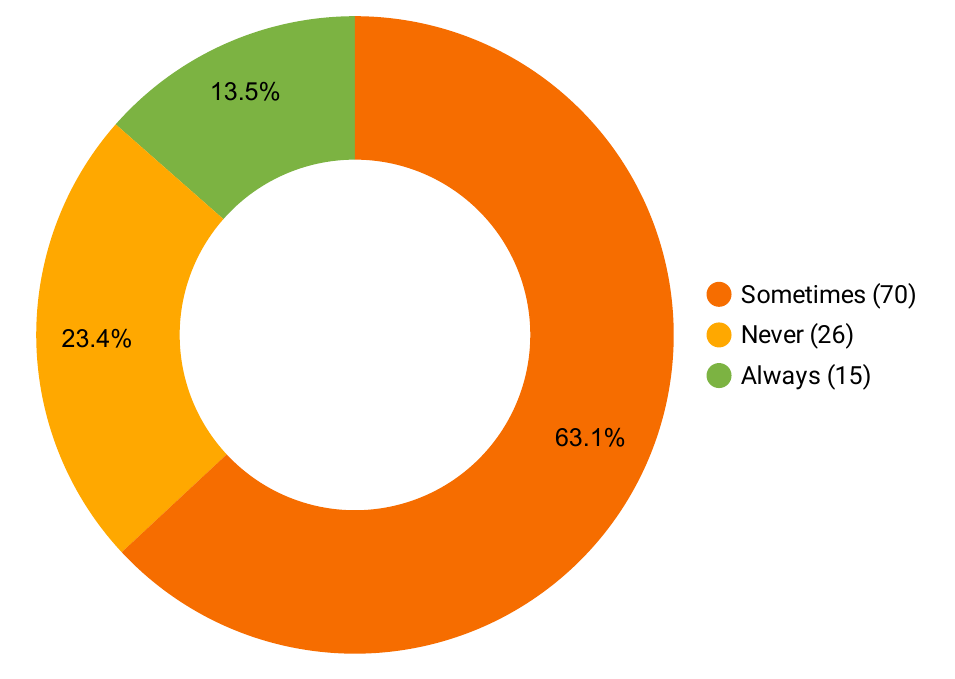}
    \caption{Distribution of participants based on how frequently they self-declare any \aigc{}}
    \label{fig:self-declaration}
    \vspace{-1em}
\end{figure}

As shown in Figure \ref{fig:self-declaration}, %the majority of developers (86.5\%) do not actively self-declare \aigc{}. Specifically, we noted that 
we see that 63.1\% of participants indicated that they \textit{sometimes} self-declare \aigc{}, 13.5\% \textit{always} self-declare, and the remaining 23.4\% \textit{never} self-declare their added \aigc{}. %Only a small percentage of participants (13.5\%) reported that they \textit{always} self-declare their added \aigc{}. 
The self-declaration practices are influenced by the extent to which \aigc{} is integrated into the project. Developers mostly tend to self-declare the use of \aigc{} when it is incorporated with little human modifications. Developers often do not feel the need to self-declare the use of \aigc{}, especially when they use \aigc{} as a guide but then significantly modify the code to meet their requirements. 

\begin{tcolorbox}[
    colback=lightgray!20, 
    colframe=darkgray,   
    boxrule=0.5mm,        
    arc=2mm,              
    title=Finding 5,      
    fonttitle=\bfseries   
]
Most participants revealed that they \textit{sometimes} self-declare \aigc{} based on the proportion of \aigc{} included in the project and the extent of human modifications to the generated code. 
\end{tcolorbox}

\subsubsection{RQ2.2: What are the reasons developers self-declare (or not) the \aigc{} in their projects?}  \label{subsubsec:reasonsSD}
We further explored the reasons behind developers' decisions to self-declare (or not) their \aigc{}. We present the key reasons we identified from the practitioners' responses.

\begin{table}[htbp]
\footnotesize
\setlength{\tabcolsep}{5pt}
\renewcommand{\arraystretch}{1.3}
\begin{tabular}{>{\raggedright\arraybackslash}p{0.18\linewidth}>{\raggedright\arraybackslash}p{0.6\linewidth}>{\raggedright\arraybackslash}p{0.13\linewidth}}
\toprule
\textbf{Category}                & \textbf{Answers to Q11 (the reasons for self-declaring AI-generated code)} & \textbf{Percentage (\%)} \\
\midrule

\multirow{2}{=}{Tracking and monitoring code for later reviews and debugging} 
& Developers use self-declaration comments to track where AI-generated code is used in the project, allowing them to separate AI-generated and human-written code. This can be helpful for future reference, quality checks, debugging, and improvements.  & \multirow{2}{*}{(43) 38.7\%} \\
\hline

\multirow{2}{=}{Transparency, accountability, and ethical considerations} 
& Developers self-declare AI-generated code for ethical reasons and to maintain transparency and accountability in the development process by clearly stating the origin of AI-generated code. & \multirow{2}{*}{(23) 20.7\%} \\
\hline

% \multirow{2}{=}{Evaluating and improving code quality} 
% & Developers self-declare AI-generated code explaining the code quality, which shows the reliability of the code and helps in code improvement in the future, thus improving code readability.  
% & \multirow{2}{*}{(15) 13.5\%} \\
% \hline

\multirow{2}{=}{Knowledge sharing within the team} 
& 
Developers self-declare AI-generated code to provide context for the code integrated into the project, which enhances knowledge sharing about development within the team and facilitates collaboration. 
& \multirow{2}{*}{(10) 9\%} \\
\bottomrule
\end{tabular}
\caption{\textcolor{black}{Categories of the reasons for self-declaring AI-generated code (Q11)}}
\label{tbl:SDReasons}
\vspace{-2em}
\end{table}

\textbf{Reasons for self-declaring AI-Generated code \aigc{}.}
Our analysis of survey responses reveals several reasons developers self-declare \aigc{}. We discuss those reasons below (a summary is provided in Table \ref{tbl:SDReasons}).

\textit{Tracking and monitoring code for later reviews and debugging.} This is the most significant cited reason, as noted by 43 (38.7\%) participants. Developers use \sdc{} to track where the added \aigc{} is used in the project, which can be helpful for future reference, quality checks, debugging, or code review. By clearly self-declaring \aigc{}, developers can more easily trace and address any issues that arise in the future due to the \aigc{}. For example, a participant explained in $R_{9}$ that self-declaration is useful for differentiating \aigc{} from human-authored code, ``\textit{by marking exactly where the \aigc{} starts and ends, it gives clear context to anyone reviewing or maintaining the code. They can easily differentiate between AI-generated and manually written parts without searching through the entire file}''. Similarly, another participant mentioned in $R_{109}$ that they ``\textit{use \aigc{} to quickly create blocks of code that would be time-consuming to write. By declaring AI use, if issues arise later, it marks it as an area of code that is worth checking first}''. \textcolor{black}{Another participant shared that they self-declare the code when it extensively integrates \aigc{} in $R_{10}$, ``\textit{I always self-declare the code when it's extensively AI-generated. It's because we need to know which part of the code is AI-generated for testing and code review purposes. But, if the code just took an idea from the AI-gen tool and then code is written by human modification as per the requirements [sic] I optionally self-declare}''. Some participants also shared that self-declaring \aigc{} is also helpful in recording the code quality as in $R_{04}$, ``\textit{It helps to keep a record of code quality and clarifies the proportion of AI-generated code versus human-written code. It also demonstrates that the code has been fully tested}''.}%These responses show that self-declaring allows the tracking of \aigc{} and helps in debugging.

\textit{Transparency, accountability, and ethical considerations.} 23 (20.7\%) participants noted reasons related to transparency, accountability, and ethics. These participants self-declare the origin of the code to maintain transparency in their teams as they believe that the ethical practice of using \aigc{} is to transparently self-declare it in the project. By including \sdc{}, developers ensure clarity on which portions of code are AI-generated, promoting responsible use of AI and, thus, promoting accountability. For instance, a participant shared in $R_{108}$ ``\textit{I self-declare \aigc{} to ensure transparency and maintain ethical standards, helping collaborators understand its origin and any unique aspects. This practice also supports clearer debugging and proper attribution}''. Another participant mentioned in $R_{12}$ ``\textit{self-declaring \aigc{} in projects promotes transparency and accountability. It helps users understand the source of the code, encourages responsible use, and allows for better collaboration between human developers and AI tools}''. \textcolor{black}{A participant particularly emphasized accountability when using \aigc{} in $R_{66}$, ``\textit{Accountability: Declaring the tool ensures that the use of AI in coding is fully documented and traceable}''.} 

% \textit{Evaluating and improving code quality.} 15 (13.5\%) participants mentioned reasons related to documenting code quality for adding \sdc{}. Added \sdc{} are used to explain code quality, indicating the reliability of the code and aiding in future code improvement. %\sdc{} also improves code readability. 
% For instance, it was noted that \sdc{} ``\textit{helps users understand the origin of the code, encourages responsible collaboration, and allows for better evaluation of the code's quality and reliability}'' ($R_{27}$). Another participant also cited how \sdc{} can provide an assessment of the code's reliability: ``\textit{it enables assessment of the code's reliability based on testing and modification history}'' ($R_{39}$). One participant indicated that self-declaring \aigc{} can be 
% Untested or partially tested code can be flagged for priority in quality assurance processes.
% %Another participant also cited how \sdc{} can help code reliability: ``\textit{self-declaring it for better readability, and in the future, if I use the same snippet, I can double-check and improve upon it}'' ($R_{94}$).  
% %These responses show that self-declaring \aigc{} improves readability and facilitates the evaluation of code quality.

\textit{Knowledge sharing within the team.} 10 (9.0\%) participants noted that they self-declare \aigc{} to encourage knowledge sharing within the team. When developers incorporate \aigc{} solutions into their projects, they help other project members understand the code by providing context for the AI-generated portion of the code. This promotes collaboration, helps other team members understand the code, and facilitates knowledge sharing among team members. A participant specifically emphasized sharing the prompts used to add context to the \aigc{}: ``\textit{including the original prompt used to generate the AI code ensures context is provided for why the code was generated in that specific way}'' ($R_{71}$). Similarly, another participant reported that ``\textit{the reason for self-declaring \aigc{} is to facilitate collaboration by providing context for other developers}'' ($R_{40}$). \textcolor{black}{A participant expressed that they self-declare \aigc{} for the understanding of other developers in the team in ($R_{72}$), ``\textit{Other developers can understand the logic or task behind the AI-generated code, making it easier to validate or modify it}''.}

\begin{tcolorbox}[
    colback=lightgray!20, 
    colframe=darkgray,   
    boxrule=0.5mm,        
    arc=2mm,              
    title=Finding 6,      
    fonttitle=\bfseries   
]
Most participants noted that they self-declare \aigc{} because of the need to track it for future code review and debugging due to concerns about code quality. Another reason that was mentioned was the promotion of the ethical use of \aigc{}. 
\end{tcolorbox}

\textbf{Reasons for \textit{not} self-declaring \aigc{}.}
Our survey responses analysis reveals several reasons explaining why developers may choose not to self-declare their \aigc{}. We summarize those reasons in Table \ref{tbl:NSDReasons}, and describe them in detail below.

\textit{Customization and modification.} One of the most frequently cited reasons, as noted by 13 (11.7\%) participants, for not self-declaring \aigc{} is that developers mostly edit and thoroughly modify the generated code according to their specific use cases. It is believed that the \aigc{} may not fully address the business requirements or may contain inaccuracies. For these reasons, developers extensively verify, modify, and test the code. %the quality and satisfy the project's requirements. 
Since the code has undergone significant changes, developers feel that it has become their own contribution. For instance, a participant noted that they thoroughly modify the \aigc{} before committing it: ``\textit{The \aigc{} that I use in my application is thoroughly understood, tested and, many times, radically changed before being committed, essentially making it original code}'' ($R_{93}$). \textcolor{black}{Another participant shared a similar idea but also expressed that they can trust the \aigc{} as their own code after their manual review and modification, ``\textit{I review code thoroughly and modify it if it's required, so it's no longer AI-generated code. I can trust it as much as I trust myself}'' ($R_{91}$).} In addition, a participant pointed out that the \aigc{} does not fulfill requirements in $R_{33}$ ``\textit{\aigc{} does not fully meet business requirements and needs to be reviewed and modified}''. \textcolor{black}{Some participants emphasized that they use \aigc{} to take an initial idea to save time, but they do not use this code blindly, as explained in the next two responses. In $R_{110}$, the developer mentioned that ``\textit{It’s irrelevant where the code came from. I don’t blindly paste code and move on, I read it, test it, and comprehend it, like any other code I would write}''. Similarly, a developer shared in $R_{03}$ that ``\textit{I only took an initial idea then modify [sic] it to meet my requirements for the problem I work on and then use it. This helps me to not search for docs and saves my time. I also add comments explaining the code if necessary}''.}
%Both responses indicate developers continuously review, change, and test \aigc{} to satisfy project requirements. Consequently, they feel that it is no longer \aigc{}, and they do not self-declare it in their projects.

\begin{table}[htbp]
\footnotesize
\setlength{\tabcolsep}{5pt}
\renewcommand{\arraystretch}{1.3}
\begin{tabular}{>{\raggedright\arraybackslash}p{0.15\linewidth}>{\raggedright\arraybackslash}p{0.62\linewidth}>{\raggedright\arraybackslash}p{0.13\linewidth}}
\toprule
\textbf{Category}  & \textbf{Answers to Q12 (reasons for not self-declaring AI-generated code)} & \textbf{Percentage (\%)} \\
\midrule

\multirow{2}{=}{Customization and modification} 
& Developers do not self-declare AI-generated code because it contains flaws and does not meet the business requirements. They thoroughly review, change, and test it before adding it to their projects, which makes it their own contribution.   
 & \multirow{2}{*}{(13) 11.7\%} \\
\hline

\multirow{2}{=}{Self-declaration is unnecessary} 
& Developers do not self-declare because they use AI code generation tools to enhance their productivity but do not completely rely on them in their development. They see it as similar to seeking help from online documentation/discussion forums (Stack Overflow) and do not feel the need to self-declare it.  & \multirow{2}{*}{(13) 11.7\%} \\
\hline

% \multirow{2}{=}{AI as a tool for increasing productivity} 
% & 5. I only took an initial idea, then modified it to meet my requirements for the problem I work on and then use it. This helps me to not search for docs and saves my time. I also add comments explaining the code if necessary. & \multirow{2}{*}{(7) 27\%} \\
% & 6. Code is a combination of my work and code gen. Don't see it as much different than copy-pasting from Stack Overflow. I've never been asked to declare code. & \\
% \hline

\multirow{2}{=}{Other reasons} 
& Developers also shared some less popular reasons, such as laziness, lack of proper requirements from the organizations, or sometimes discouragement from organizations due to security and legitimacy concerns with AI-generated code.  & \multirow{2}{*}{(6) 5.4\%} \\
\bottomrule

\end{tabular}
\caption{Categories of the reasons for not self-declaring AI-generated code (Q12)}
\label{tbl:NSDReasons}
\vspace{-2em}
\end{table}

\textit{Self-declaration is unnecessary.} The other top-cited reason, as mentioned by 13 (11.7\%) participants, for not self-declaring \aigc{} is that some developers see it as an unnecessary activity. It was noted that developers may utilize AI code generation tools to generate ideas; however, they emphasized that they only employ them to enhance their productivity. They see \aigc{} as similar to seeking help from online documentation or developer forums (such as Stack Overflow). Those AI code generation tools are seen as assisting tools. Hence, they believe that self-declaration of AI-generated code is an unnecessary activity. For example, a participant answered that: ``\textit{I usually search for small code snippets in languages I don't remember the syntax for, and I believe that adding comments specifying the source would be unnecessary for my team}'' ($R_{95}$). 
Another participant noted the following: ``\textit{In the rare cases where I'm not providing my own original code, it's going through multiple iterations of my prompting and manually editing to get what I need ..... I don't self-declare using a spell-checker, so I don't see the value in self-declaring \aigc{}}''. \textcolor{black}{A participant expressed that they use AI code generation tools as an assisting tool but not as an autonomous contributor in $R_{92}$, ``\textit{I use Github Copilot, which writes code alongside me. When I am doing any functionality, it keeps giving me suggestions and one of them suits me so I just select it and then after making desirable changes, it's good to go. Since it's combination of both me and Copilot writing code, I don't feel the need of specifying something}''. Some participants mentioned that they see \aigc{} similar to finding solutions from code documentation or developer forums, as in $R_{87}$ ``\textit{No need [for self-declaring \aigc{}] because it would be like declaring some code that comes from documentation or StackOverflow or a colleagues' [sic] advice}''.}

\begin{tcolorbox}[
    colback=lightgray!20, 
    colframe=darkgray,   
    boxrule=0.5mm,        
    arc=2mm,              
    title=Finding 7,      
    fonttitle=\bfseries
]
Participants noted that they do not self-declare \aigc{} because the code requires human review and modification before integration into their projects, and they feel that self-declaration is not required, as it is similar to using online documentation or tools like a spell checker.  
\end{tcolorbox}

\section{Discussion} \label{sec:discussion}

\subsection{Developers Practices in Declaring AI-Generated Code}
% Our study findings demonstrate that developers follow different ways to self-declare \aigc{} in their projects. Developers include varying levels of detail in the \sdc{}, ranging from simple one-liners to detailed explanations and contextual information about the added code. Developers' approaches to the scope of \sdc{} also vary - sometimes self-declaring specific code snippets or entire code files as AI-generated. Developers utilize these ways not only to provide simple annotations but also to address ethical, collaborative, and quality-related concerns. 
In this section, we discuss the key reasons why developers use various ways to self-declare \aigc{}.

\subsubsection{\textcolor{black}{Developers Mostly Use Simple Self-Declaration Comments.}}
\textcolor{black}{In our mining study, the highest number of \sdc{} were simple one-line comments, indicating that the code was generated by an AI code generation tool. This finding aligns with our practitioners’ survey results, where the top reasons for self-declaring \aigc{} were tracking and monitoring AI-generated code and fulfilling ethical or transparency requirements. These goals can be effectively met using simple self-declaration comments, which allow developers to mark \aigc{} without adding additional information that may make self-declaration inefficient. Therefore, these results from the mining study and the practitioners’ survey complement each other: simple self-declaration comments enable both accurate traceability and ethical acknowledgment of AI use in a practical and straightforward manner. These findings align with existing studies, such as Yu \textit{et al.}, who also found that comments acknowledging the use of AI are mostly short and lack details \cite{yu2024large}. Additionally, simple self-declaration comments help developers maintain transparency, preserve the provenance of generated code, and integrate \aigc{} efficiently without adding unnecessary complexity to their workflows.}

\subsubsection{\textcolor{black}{Developers Attach Additional Information in Self-Declaration Comments.}}
\textcolor{black}{Our mining study findings show that developers also enrich self-declaration comments with contextual information, such as prompts, code explanations, and code quality indicators. Developers utilize these ways not only to provide simple annotations but also to address ethical, collaborative, and quality-related concerns on AI-generated code. These ways of self-declaring \aigc{} have various benefits for developers. (1) Self-declaration comments can contain code quality indications sharing details about the quality and the reliability of the integrated \aigc{}, which is also highlighted in existing literature~\cite{fu2023security, tambon2025bugs}. (2) Self-declaration comments can also contain contextual information or code explanations explaining the context and rationale of integrated \aigc{}. This practice is also aligned with existing studies, such as the survey by Stalnaker \textit{et al.}, which shows that developers record the prompts or note the whole interaction with AI tools to document \aigc{} \cite{stalnaker2024developer}. Our survey participants also supported these ways of writing \sdc{} (see Figure \ref{fig:practices}). However, the challenge with attaching additional information is to maintain a balance in the amount of details to be shared. This is because overly detailed comments may hinder readability, while insufficient detail may fail to provide meaningful information about the added \aigc{}.} 
% Similarly, Yu \textit{et al.} reported that developers share the prompts used to generate code \cite{yu2024large}. 

\subsubsection{Snippet-level vs. File-level self-Declaration Comments.}
% We categorized the scope of \sdc{} into two categories: \textit{\snl{}} and \textit{\fl{}} \sdc{}. 
% \textit{File-level comments.} When developers add multiple \aigc{} snippets in one source file, or the source file is mainly written with assistance from AI code generation tools, they provide a single self-declaration comment at the start or end of the source file to acknowledge that the file contains \aigc{}. %The \fl{} comments apply to the whole source file, possibly meaning that the entire file is AI-generated. 
% The drawback of \fl{} comments is that they are provided at a high level and can not help accurately track the \aigc{} (i.e., which part of the file is AI-generated).

% \textit{Snippet-level comments.} When developers use short \aigc{} snippets in their projects, they mention precisely which specific segments are AI-generated. %They tend to self-declare whole classes or functions. In those cases, 
% The \sdc{} only apply to the specific scope of the self-declared AI-generated code snippets. For example, if a developer writes a \snl{} comment above a function, then the scope of this comment applies only to that specific function. The main benefit of \snl{} self-declaration is that it accurately describes which code snippet is AI-generated, allowing for the separation of AI-generated from the human-written code within a code file. This helps in tracking the \aigc{} in later code reviews, bug fixing, and quality improvement processes.
The preference between \snl{} and \fl{} \sdc{} reflects developers' trade-offs between granularity and efficiency. Developers prefer \snl{} \sdc{} when they need to track and differentiate \aigc{} accurately. On the other hand, developers use \fl{} \sdc{} when the code file contains code that is predominantly AI-generated. File-level comments are efficient as they save time and effort, but sacrifice granularity, making it difficult to trace individual \aigc{} snippets within code files. Developers' preference for \snl{} comments, as indicated by the practitioners’ survey and mining study, demonstrates that they prioritize accurate traceability and accountability over efficiency when integrating \aigc{} in their projects. 

\subsubsection{\textcolor{black}{Most Repositories Contain Only a Single File With Self-Declared AI-Generated Code.}}
\textcolor{black}{While 69\% of repositories in the mining study contained only one file with self-declared AI-generated code, our survey found that 63\% of developers ``\textit{sometimes}'' self-declare \aigc{}. At first glance, this might appear inconsistent, but a closer look clarifies the difference. The survey results indicate that these developers self-declare \aigc{} selectively, in cases when the code has minimal human intervention or has not been properly tested. It can be explained using a participant's response in $R_{51}$, ``\textit{I rarely use it in some part of my project so I only declare when it is used more than 30\% in my project}''. Only 13.5\% of developers reported ``\textit{always}'' self-declaring AI-generated code. This selective behavior explains why repositories often contain only a single file with such comments: the majority of developers self-declare \aigc{} only in selective cases when they consider necessary.}

\textcolor{black}{Furthermore, the survey results suggest that developers are generally willing to self-declare AI-generated code, with 63\% indicating that they self-declare \aigc{} at least ``\textit{sometimes}'', and only 23\% reporting that they never self-declare \aigc{}. This willingness implies that with better tool support, such as automated mechanisms for detecting AI-generated code and helping developers insert self-declaration comments, the adoption of the practices of self-declaring \aigc{} could increase significantly. This is likely because these tools would reduce the perceived effort and liability associated with self-declaring \aigc{}.}

\subsubsection{Distribution of Self-Declared AI-Generated Code in GitHub Repositories.}
The distribution of self-declared \aigc{} across repositories (shown in Figure~\ref{fig:distribution}) suggests two possible interpretations: (1) Even with the introduction of AI code generation tools in software development, developers still use these tools in a limited manner and do not rely on them due to code quality \cite{Nikolaidis2023, dakhel2023github} and security concerns \cite{fu2023security}. The low adoption rate also indicates that developers are still exploring the potential of AI tools and adding them selectively into their development workflows. (2) The actual distribution of \aigc{} in these projects may be higher than reported, as our analysis relies solely on self-declaration of AI-generated code by the developers. Some developers may integrate \aigc{} without declaration or acknowledgment. In our survey, some participants mentioned that they add \aigc{} without self-declaration (see the discussion in Section~\ref{notsdreason}).  

% \subsubsection{Developer Satisfaction with AI-Generated Code.}
% By analyzing developers' sentiments expressed in their self-declaration comments, we found that most developers were satisfied and reported a positive experience in generating the desired AI-generated code. 
% % They reported a positive experience in generating their desired code. 
% % By analyzing the sentiments of the \sdc{}, we found that most developers are positive about their \aigc{} as they found it helpful and were positive about their experience with the \aigc{} in generating their desired code. 
% These results align with previous findings, which show that AI code generation tools enhance developers' productivity \cite{Peng2023The} and are useful in coding tasks \cite{Nikolaidis2023}. These tools enable developers to focus on higher-level problem-solving tasks, as they reduce the time and effort required for development. However, this satisfaction is related to the ability to control and refine the outputs of code generation tools, as highlighted by the emphasis on iterative prompting and human intervention \cite{tony2024prompting, chen2023unleashing}. 
% % This finding emphasizes the significance of developing AI tools that are not only effective but also controllable and usable, enabling developers to accomplish their expected outputs effectively. 
% % Furthermore, AI tools increase developers' productivity, which in turn impacts satisfaction \cite{storey2019towards}.

\subsection{Reasons for Self-Declaring (or Not) AI-Generated Code}
Developers shared mixed opinions about their reasons for self-declaring their \aigc{}. We discuss those reasons below. 

\subsubsection{Reasons for Self-Declaring AI-Generated Code.}
% Our survey results show that developers self-declare \aigc {} due to several reasons, including (1) the need for tracking and monitoring \aigc{} in code files, (2) promoting ethical use of \aigc{}, (3) providing context to share knowledge about \aigc{} and enhance collaboration, and (4) indicating the code quality by sharing whether the code is tested or undergone human modification. 

\textcolor{black}{Our survey shows that the primary reasons developers self-declare \aigc{} are the need for tracking and monitoring AI-generated code, as well as ethical considerations. Existing literature has extensively highlighted the quality and security risks associated with AI-generated code, emphasizing the importance of reviewing and improving it by experienced human developers \cite{jin2024can, ouyang2024empirical}. This highlights the need for accurately tracking AI-generated contributions to support future code review, debugging, and maintenance. While several studies have explored automated techniques for detecting AI-generated code \cite{gurioli2025isthisyou,li2023discriminating}, these approaches face challenges in identifying \aigc{} accurately. Interestingly, developers also shared in their responses that they integrate \aigc{} after modifying and improving it. This modification can further reduce the accuracy when detecting \aigc{} using automatic approaches. This highlights the importance of self-declaring \aigc{} for accurately tracking it as suggested by survey participants. Additionally, developers considered self-declaring \aigc{} as an ethical obligation. This shows developers' concern for ethical AI-assisted software development and responsible human-AI collaboration. Developers noted that self-declaring \aigc{} facilitates transparency in the development process and helps track its provenance. Several previous studies have explored ethics in AI, specifically in software development \cite{stalnaker2024developer,xu2024distinguishing}. Nonetheless, existing research lacks clear guidelines on the ethical use of \aigc{} in practice.}

\subsubsection{Reasons for \textit{Not} Self-Declaring AI-Generated Code.} \label{notsdreason} 
% Our survey results show that there are two reasons for not self-declaring \aigc {}: (1) the extensive modifications to \aigc{} and (2) the perception of AI code generation tools as productivity enhancers rather than autonomous contributors. 
\textcolor{black}{The main reason is that developers do not self-declare \aigc{} because the code undergoes a cycle of iterative refinement before it is added to their projects. The rationale behind this modification is twofold: (1) Developers use the generated code as a source of inspiration only. Although AI code generation tools have undergone significant improvements over the years, their output often still requires customization to meet project requirements. (2) AI-generated code may contain code quality issues that require human modification before it becomes a viable solution. As developers extensively verify and modify the code according to their requirements, the generated code evolves through additional human contributions. Therefore, developers feel that the code is now their contribution. Another reason is that developers view these AI code generation tools similarly to reference materials, such as online documentation and coding forums (e.g., Stack Overflow). Therefore, they view the generated code as similar to adapting examples from online forums or documentation. Developers utilize AI code generation tools to enhance productivity; however, they do not employ these tools as autonomous contributors. Therefore, they do not feel the need to self-declare \aigc{}.}

\subsection{Implications}\label{guidelines}

\subsubsection{\textcolor{black}{Implications for Researchers}}
\textcolor{black}{Our findings highlight several implications for the future of AI-assisted software engineering. (1) The low frequency of self-declaration comments suggests the need for automated tools that can detect AI-generated code and insert structured self-declaration comments with the developer's confirmation to reduce manual effort while improving traceability. Such tools could also allow the attachment of useful additional information, such as prompts and quality indications, to support debugging and code review. (2) The use of snippet-level and file-level comments observed in our study indicates that future IDEs should provide flexible mechanisms for comments to self-declare AI-generated contributions at different levels of granularity. (3) The inclusion of quality indicators in self-declaration comments shows potential for integrating automated quality checks with self-declaration comments (e.g., attaching a negative indicator to a self-declaration comment if unit tests or static analysis fail). This feature will enable teams to better assess the reliability of AI-generated code. (4) Developers’ ethical motivations for self-declaration indicate the importance of establishing organizational guidelines and industry standards for \aigc{} provenance, while also creating opportunities for future research to explore how automated self-declaration can be integrated into existing development workflows.}

\subsubsection{\textcolor{black}{Implications for Practitioners}}
Based on our results, we offer a set of guidelines for developers regarding how and when to declare \aigc{}. This includes recommendations on what information developers should include in their \sdc{} and how to effectively manage the scope of \sdc{}.

\textit{Self-declaring \aigc{}, especially when the generated code is not fully understood or tested.} Developers should self-declare \aigc{}, particularly when it has been generated with little human intervention, or when the generated code has not been thoroughly tested or verified. Even though the code is thoroughly reviewed, understood, and tested, self-declaration remains valuable for the ethical use of \aigc{}. \textcolor{black}{\textbf{\textit{Relevance.}} Our survey revealed that developers often self-declare AI-generated code mainly to support later debugging and review, driven by quality and security concerns. Even when code is modified or tested, many developers view self-declaration as an ethical practice that ensures transparency.}

\textit{Using simple self-declaration comments for better transparency.} When adding AI-generated code to a project, developers may include a short self-declaration comment. This comment should note that the code was AI-generated, the name of the tool used, and the name of the code snippet (e.g., the name of the class, function, or language construct). \textcolor{black}{\textbf{\textit{Relevance.}} Our mining study showed that developers commonly include short self-declaration comments, and developers in the practitioners’ survey shared that self-declaring AI-generated code helps ensure transparency in development, promotes the ethical use of AI-generated code, and facilitates the distinction between AI-generated and human-written code.} 

\textit{Adding code explanations for complex AI-generated code.} A good practice for developers is to include a brief explanation alongside the simple \sdc{}, especially when the \aigc{} is relatively complex. These comments should describe how the code works, why it is added, and any key details about the integrated \aigc{} that team members may need in the future. These code explanations should be kept concise, and developers should be mindful of the code's readability. \textcolor{black}{\textbf{\textit{Relevance.}} Our mining study revealed that developers frequently include code explanations alongside self-declaration comments. Survey participants supported this practice, especially when the generated code is complex, as they believe it improves code maintainability and enhances knowledge sharing within the team.}

\textit{Include code quality indications in \sdc{}.} Developers should add details about the quality of \aigc{} in the self-declaration comments. This can include whether the code has been tested, whether it was modified by a human, and if there are any known issues or limitations in the \aigc{}. Including such information helps team members assess the code's reliability and determine if further improvements are necessary. \textcolor{black}{\textbf{\textit{Relevance.}} In the mining study, we observed that developers often add comments about code quality, especially regarding whether AI-generated code has been tested or has known limitations. Additionally, survey participants also highlighted that quality concerns are a key reason for self-declaring AI-generated code.}

\textit{Document the context of \aigc{}.} When using AI-generated code, developers should consider documenting its use context, such as the prompts used. Keeping a record of such information can help developers understand how the code was generated and make it easier to modify or troubleshoot in the future. \textcolor{black}{\textbf{\textit{Relevance.}} Our mining study identified several instances where developers included contextual information, such as prompts used during code generation.}

\textit{Using snippet-level comments for better traceability.} It is advised to use \snl{} rather than \fl{} comments when self-declaring AI-generated code, as they provide a precise and more detailed approach to track \aigc{} snippets. However, if a file contains mostly AI-generated code or multiple \aigc{} snippets, \fl{} comments may be more suitable, as they offer a more concise summary without excessive repetition. \textcolor{black}{\textbf{\textit{Relevance.}} The mining study and survey responses revealed that developers predominantly use snippet-level comments for precise tracking of AI-generated code, but utilize file-level comments for efficiency of self-declaration when the file is mostly AI-generated.}

\section{Threats to Validity} \label{sec:validity}

\subsection{Internal Validity}
\textit{Possible bias in manual analysis.} During the data collection of the mining study, we first searched for \aigc{} on GitHub repositories using a variety of search terms. After we got the results, we manually labeled the code snippets as AI-generated. This approach may introduce bias in our manual labeling of \aigc{} snippets. To reduce this bias, the first author performed the labeling, and then the other co-authors cross-validated the labeling results. In cases of disagreements, we performed the labeling process through extensive discussion and continually measured the agreement level (Cohen's Kappa coefficient) among the authors. 

\textit{Participant familiarity with \aigc{}.} Our study focuses specifically on \aigc{}, and we wanted survey participants who actively use AI code generation tools in their projects. This requirement introduces a threat that some respondents may lack experience with AI code generation tools or may prefer not to use them in their software projects. To minimize this risk, we included a prerequisite question on the first page of the survey that asks participants if they use \aigc{} in their software projects. Based on their answers, we excluded 12 responses from participants who did not use \aigc{} in their software projects. We also clarified that we invite participants who utilize \aigc{} in their projects on the first page of our survey questionnaire.

\textit{Selection of participants.} We identified survey participants from GitHub repository contributors, our personal contacts, and developer groups, and invited them via email and social media platforms. A potential threat to validity arises if some participants lack experience in software development, making them ineligible to participate in our survey. To minimize this threat, we explicitly mentioned on the first page of our survey that we invite participants who are professionally working in the software industry. In addition, we also included questions about education (survey question Q3) and experience (survey question Q4) in the demographic questions, helping us to ensure that we only received responses from participants with the appropriate software development background.

\subsection{Construct Validity}
\textit{Keyword-based search.} We used a set of search terms to identify and locate \aigc{} on GitHub repositories. This approach introduced a threat to validity as the search terms used might not be comprehensive or inclusive. % in collecting \aigc{} snippets using varied ways to self-declare \aigc{}. 
To minimize this threat, we first conducted a pilot search using four search strategies and tested their corresponding groups of search terms (see Table 1), which helped us finalize the list of potential search terms. We iteratively added new terms to improve the coverage of our keyword. As a result, we identified a list of 23 search terms with broad coverage. %that helped us to collect a broader range of data for self-declaration of \aigc{} and identified different ways developers self-declare \aigc{}. 
\textcolor{black}{The finalized 23 search terms include \textit{thanking} terms, such as ``\textit{Thanks ChatGPT}''. These search terms introduced a threat to the validity because they could introduce bias in the study results. To minimize this threat in data collection, we also included \textit{warning} search terms such as ``\textit{warning Copilot}'' and also searched using these terms in addition to \textit{thanking} terms (see Section \ref{subsubsec:keyword}). Secondly, we also ensured that all self-declaration comments appreciating AI tools are not included using \textit{thanking} terms. Consequently, we have a large number of self-declaration comments appreciating AI tools, but they were not collected using \textit{thanking} search terms.}

\textit{Understandability of the survey questionnaire.} We created our survey questionnaire with a total of 12 questions, including nine closed-ended and three open-ended questions. The survey questions may introduce a threat to construct validity. To overcome this threat, we employed two strategies: (1) We requested the assistance of an experienced software engineer researcher to review the survey questionnaire, which helped us refine the clarity of the survey questions and the structure of the questionnaire. (2) We conducted a pilot run of the survey with a small subset of our target survey population. We asked the participants in the pilot survey to provide feedback on the clarity of the survey. We used this feedback to improve the readability and understandability of our survey questionnaire.

%\textit{Closed-ended survey questions.} Our survey questionnaire contains nine closed-ended questions. This raises the threat of insufficient options available for the participants to answer these questions. In order to minimize this threat, the options to answer the closed-ended questions were provided based on the results of the mining study. In addition, we also included an open text area so that the participants could share their ideas in addition to the provided options in the closed-ended questions. Consequently, we received responses where the participants shared their answers in the open text areas besides the provided options in the closed-ended questions.

\subsection{External Validity}
 \textit{Generalizability of the survey results.} Our survey received 111 valid responses from practitioners. Although the survey provides important information about the practices of self-declaring \aigc{}, the sample of participants can be viewed as not fully representative. %The findings from this sample may not be applicable to other countries and regions. 
To minimize this threat, we aimed to collect responses from diverse countries and regions to improve the generalizability of our findings. In addition, the diversity of the survey participants, including their experience, roles within the current organization, and application domains (see Section~\ref{subsec:surveydemographics}), partially mitigates this threat.

\textit{Collecting code snippets only from GitHub.} We collected \aigc{} snippets from GitHub repositories. Considering that \aigc{} is also available on other coding platforms, this poses a threat to the external validity of our study results. To minimize this threat, we compiled a dataset of 613 code snippets from repositories across the top 10 programming languages, ensuring a broad coverage. Given GitHub's position as the leading open-source data source and its common use in similar research, this threat is reasonably mitigated.

\subsection{Reliability}
The reliability of empirical studies refers to the ability to replicate results and findings under similar conditions. A potential threat to reliability arises from the manual collection and analysis of \aigc{} snippets from GitHub projects and the qualitative analysis of the survey responses. To mitigate this threat, we followed a rigorous process for collecting and analyzing data from both sources (see Section \ref{subsec:Miningstudydesign} and Section \ref{subsec:surveydesign}).
% We followed a strict filtration process in which all authors were involved, and the results were regularly cross-validated. 
We also made the dataset used in the study \cite{replicationPackage} available for other researchers to replicate our findings.

\section{Conclusions and Future Work} \label{sec:conclusion}
This study explores the practices of developers' self-declaration of \aigc{}. We examined whether developers self-declare \aigc{}, their motivations for doing so, and the methods they use to achieve this. We employed a mixed-methods approach, combining data from GitHub repositories with a follow-up practitioners’ survey among practitioners. We found that developers have varying opinions on self-declaring \aigc{}, and their self-declaration practices are influenced by the proportion of AI contribution in the generated code. Most practitioners (76.6\%) \textit{always} or \textit{sometimes} self-declare \aigc{}. In contrast, other practitioners (23.4\%) noted that they \textit{never} self-declare \aigc{}. Top reasons for self-declaring \aigc{} include (1) the need to track and monitor code for future review and debugging, and (2) ethical considerations related to the use of code generation tools. On the other hand, reasons for not self-declaring \aigc{} include (1) AI-generated code often requires extensive human modification, and (2) developers view self-declaration as an unnecessary activity, comparable to utilizing online documentation or coding forums. 
We then provided implications for researchers and a set of guidelines for developers to follow when self-declaring \aigc{}. %We encourage further exploration of this topic, particularly with varied datasets and in different development environments.
%\subsection{Future Work}
%We studied the practice of self-declaring \aigc{} among developers to understand the motivations behind such declarations. 

Our study also opens up several directions for future research. (1) We identify various reasons for self-declaring (or not) \aigc{}, which can be further investigated with a larger set of developers. Additionally, we provide guidelines for self-declaring \aigc{} by analyzing how developers report their generated code based on data from the mining study and practitioners’ survey. Future research can further refine and expand the guidelines using new datasets and validate their usefulness through empirical evaluation. (2) Our research shows that developers consider \sdc{} beneficial for tracking \aigc{} for future review, debugging, and improvement. This opens up avenues for future research to investigate the impact of \sdc{} on various code quality attributes, such as readability and maintainability. Such studies can contribute to a deeper understanding of the evolving landscape of software development, where \aigc{} is increasingly integrated into modern software systems. %These research directions could contribute significantly to the software engineering community.
(3) It would be valuable to provide a code-commenting function in AI code generation tools to support the guidelines (see Section~\ref{guidelines}) for developers to declare \aigc{}. \textcolor{black}{(4) Our study focuses on the self-declaration of code snippets generated by AI code generation tools. With the growing popularity of autonomous coding agents capable of generating entire code files, future research can investigate the self-declaration of AI-generated code in the context of these coding agents.}

\section*{Data Availability}
The dataset used in this work is available at \cite{replicationPackage}.

\begin{acks}
This work has been partially supported by the National Natural Science Foundation of China (NSFC) with Grant No. 62172311 and the Major Science and Technology Project of Hubei Province under Grant No. 2024BAA008. The numerical calculations in this paper have been done on the supercomputing system in the Supercomputing Center of Wuhan University.
\end{acks}

\end{sloppypar}

\bibliographystyle{ACM-Reference-Format}
\bibliography{basebib}

%%
% %% If your work has an appendix, this is the place to put it.
% \appendix









\end{document}